%% file: o4b-ssm.tex
\documentclass[superscriptaddress,twocolumn,nopreprintnumbers,floatfix,nofootinbib,table]{aastex631}
\pdfoutput=1

\newif\ifprintauthors
\printauthorstrue

\usepackage{amsmath}
\let\tablenum\relax

\usepackage[T1]{fontenc}
\usepackage[utf8]{inputenc}
\usepackage{amssymb,graphicx}
\usepackage{scrextend}
\usepackage[caption=false]{subfig}

\usepackage{multirow}

\usepackage{siunitx}
\DeclareSIUnit\parsec{pc}
\DeclareSIUnit\Mpc{\mega\parsec}
\DeclareSIUnit\yr{yr}

\usepackage{txfonts}
\newlength{\capheight}
\usepackage{textgreek}

\usepackage{xspace}
\usepackage{array}
\usepackage{booktabs}
\usepackage{color}
\usepackage{xcolor}
\usepackage{colorblind}
\usepackage{makecell}

\usepackage{acronym}

\usepackage{orcidlink}

\input{macros}
\input{macros_from_json}
\input{units_symbols}

\input{gwtc-common-files__acronyms}
\input{gwtc-common-files__commands}

\input{gwtc-common-files__macros__symbol_macros}

\input{gwtc-common-files__macros__misc_macros}

\input{gwtc-common-files__macros__datetime_macros}

\input{gwtc-common-files__macros__quantities_macros}

\input{acronyms}

\definecolor{CBcyan}{RGB}{0,114,178}
\definecolor{CBdarkorange}{RGB}{230,159,0}
\definecolor{CBgreen}{RGB}{0,158,115}
\definecolor{CBviolet}{RGB}{204,121,167}
\definecolor{CBSkyBlue}{RGB}{86,180,233}
\definecolor{CBVermilion}{RGB}{213,94,0}
\definecolor{CByellow}{RGB}{240,228,66}
\definecolor{CBGray}{RGB}{128,128,128}
\definecolor{CBBlack}{RGB}{0,0,0}

\definecolor{BlueGreen}{RGB}{0, 128, 128} 
\definecolor{Red}{RGB}{255, 0, 0} 
\definecolor{Green}{RGB}{0, 128, 0}
\definecolor{Pink}{RGB}{255, 0, 255}

\newcommand{\reviewed}[1]{#1}

\definecolor{REVIEWED}{rgb}{0,0,0}

\input{results_numbers_macros}

\begin{document}

\title{Search for Compact Binary Coalescences with Sub-solar Mass Components in Data from the Second Part of LIGO--Virgo--KAGRA's Fourth Observing Run} 

\ifprintauthors
\input{authorlist__LSC-Virgo-KAGRA-Authors-Feb-2026-aas-fragment.tex}
\fi
\collaboration{0}{The LIGO Scientific Collaboration, the Virgo Collaboration, and the KAGRA Collaboration}

\correspondingauthor{LSC P\&P Committee, via LVK Publications as proxy}
\email{lvc.publications@ligo.org}

\begin{abstract}
\input{source__abstract}
\bigskip
\noindent

\end{abstract}


\input{source__1-introduction}

\input{source__2-search}

\input{source__3-search_sensitivity}

\input{source__4-rates}

\input{source__5-dm_constraints}

\input{source__6-conclusion}

\section*{Data availability}
\input{data_availability.tex}

\section*{Acknowledgements}
\input{LVK_P2000488_v31.tex}

\facility{LIGO}
\facility{Virgo}
\software{\input{software_ack.tex}}

\clearpage
\bibliography{}

\ifprintauthors
\allauthors
\fi
\end{document}

%% file: macros.tex
\NewDocumentCommand{\numRound}{O{2} m}{\num[round-mode=figures, round-precision=#1]{#2}}
\NewDocumentCommand{\numRoundSci}{O{2} m}{\num[scientific-notation=true, round-mode=figures, round-precision=#1]{#2}}

\newcommand{\gwtcfour}{\gwtc[4.0]\xspace}
\newcommand{\gwtcfive}{{\thisgwtc}\xspace}
\newcommand{\Ofourab}{O4a+b\xspace}

%% file: units_symbols.tex
\newcommand{\spincomponent}{\ensuremath{\chi_{1,2}}\xspace}

\newcommand{\masscomponent}{\ensuremath{m_{1,2}}\xspace}

%% file: gwtc-common-files__acronyms.tex
\usepackage{amstext}

\makeatletter
\@ifpackageloaded{cleveref}{
  \AtBeginDocument{
    \def\ltx@label#1{\cref@label{#1}}%
    \def\label@in@display@noarg#1{\cref@old@label@in@display{#1}}%
    \def\label@in@mmeasure@noarg#1{%
      \begingroup%
        \measuring@false%
        \cref@old@label@in@display{#1}%
      \endgroup}%
  }
}{}
\makeatother

\protected\def\protectedacused{\acused}

\acrodef{LIGO}[LIGO]{Laser Interferometer Gravitational-Wave Observatory}
\acrodef{LHO}[LHO]{\ac{LIGO} Hanford Observatory}
\acrodef{LLO}[LLO]{\ac{LIGO} Livingston Observatory}
\acrodef{KAGRA}[KAGRA]{KAGRA}\acused{KAGRA}
\acrodef{iKAGRA}[iKAGRA]{initial \ac{KAGRA}}
\acrodef{bKAGRA}[bKAGRA]{baseline \ac{KAGRA}}
\acrodef{GEO}[GEO\,600]{GEO\,600 \ac{GW} detector}
\acrodef{aLIGO}{Advanced \ac{LIGO}}
\acrodef{A+}{Advanced+ \ac{LIGO}}
\acrodef{Asharp}[\ensuremath{\text{A}^\sharp}]{\ac{LIGO} \acs{Asharp}}
\acrodef{AdV}{Advanced \acl{Virgo}}
\acrodef{AdV+}{Advanced \acl{Virgo}+}
\acrodef{Virgo}{Virgo}\acused{Virgo}
\acrodef{VirgoNEXT}[Virgo\_nEXT]{Virgo\_nEXT}\acused{VirgoNEXT}

\acrodef{LSC}[LSC]{\acs{LIGO} Scientific Collaboration}
\acrodef{LV}[LV]{\acs{LIGO}--\acs{Virgo} Collaboration\protect\protectedacused{LVC}}
\acrodef{LVC}[LV]{\acs{LIGO}--\acs{Virgo} Collaboration\protect\protectedacused{LV}}
\acrodef{LVK}[LVK]{\acs{LIGO}--\ac{Virgo}--\ac{KAGRA} Collaboration}
\acrodef{IGWN}[IGWN]{International \ac{GWH} Observatory Network}

\acrodef{O1}[O1]{first observing run}
\acrodef{O2}[O2]{second observing run}
\acrodef{O3}[O3]{third observing run}
\acrodef{O3a}[O3a]{first half of the third observing run}
\acrodef{O3b}[O3b]{second half of the third observing run}
\acrodef{O3GK}[O3GK]{observing run}
\acrodef{O4}[O4]{fourth observing run}
\acrodef{O4a}[O4a]{first part of the fourth observing run}
\acrodef{O4b}[O4b]{second part of the fourth observing run}
\acrodef{O4c}[O4c]{third part of the fourth observing run}
\acrodef{IR1}[IR1]{intermediate run 1}
\acrodef{O5}[O5]{fifth observing run}

\acrodef{BH}[BH]{black hole}
\acrodef{BBH}[BBH]{binary \ac{BH}}
\acrodef{BNS}[BNS]{binary \ac{NS}}
\acrodef{IMBH}[IMBH]{intermediate-mass \ac{BH}}
\acrodef{NS}[NS]{neutron star}
\acrodef{BHNS}[BHNS]{\ac{BH}--\ac{NS} binary}
\acrodefplural{BHNS}[BHNSs]{\ac{BH}--\ac{NS} binaries}
\acrodef{NSBH}[NSBH]{\ac{NS}--\ac{BH} binary}
\acrodefplural{NSBH}[NSBHs]{\ac{NS}--\ac{BH} binaries}
\acrodef{PBH}[PBH]{primordial \ac{BH}}
\acrodef{CBC}[CBC]{compact binary coalescence}

\acrodef{GW}[GW]{gravitational wave\protect\protectedacused{GWH}}
\acrodef{GWH}[GW]{gravitational-wave\protect\protectedacused{GW}}
\acrodef{IFO}[IFO]{interferometer}
\acrodef{SNR}[SNR]{signal-to-noise ratio}
\acrodef{FAR}[FAR]{false-alarm rate}
\acrodef{IFAR}[IFAR]{inverse false-alarm rate}
\acrodef{FAP}[FAP]{false alarm probability}
\acrodef{PSD}[PSD]{power spectral density}
\acrodef{GR}[GR]{general relativity}
\acrodef{NR}[NR]{numerical relativity}
\acrodef{PN}[PN]{post-Newtonian}
\acrodef{EOB}[EOB]{effective-one-body}
\acrodef{ROM}[ROM]{reduced-order model}
\acrodef{IMR}[IMR]{inspiral--merger--ringdown}
\acrodef{PDF}[pdf]{probability density function}
\acrodef{PE}[PE]{parameter estimation}
\acrodef{CI}[CI]{credible interval}
\acrodef{CL}[CL]{credible level}
\acrodef{EOS}[EoS]{equation of state}
\acrodef{KLD}[KLD]{Kullback--Leibler divergence}
\acrodef{JSD}[JSD]{Jensen--Shannon divergence}
\acrodef{GCN}[GCN]{General Coordinates Network}
\acrodef{GWTC}[GWTC]{Gravitational-Wave Transient Catalog}
\acrodef{GWOSC}[GWOSC]{Gravitational Wave Open Science Center}
\acrodef{WDM}[WDM]{Wilson--Daubechies--Meyer}
\acrodef{DQR}[DQR]{data-quality report}

\acrodef{CWB}[cWB]{coherent WaveBurst}
\acrodef{LAL}[LAL]{\ac{LIGO} algorithm library}

\acrodef{CHRoCC}{central heating radius of curvature correction}
\acrodef{NonSENS}{non-stationary estimation and noise subtraction}
\acrodef{RF}{radio frequency}
\acrodef{PNC}{phase noise cancellation}
\acrodef{ASC}{alignment sensing and control}
\acrodef{WFS}{wave-front sensing}
\acrodef{BPC}{beam position control}
\acrodef{ADS}{alignment dither systems}
\acrodef{OMC}{output mode cleaner}
\acrodef{LVDTs}{linear variable differential transformers}
\acrodef{GAS}{geometrical anti-spring}

\acrodef{PTA}{Pulsar Timing Array}

%% file: gwtc-common-files__commands.tex
\newcommand\gwtc[1][?]{\mbox{GWTC\if#1?\else-#1\fi}}
\newcommand\thisgwtcversionmajor{5}
\newcommand\thisgwtcversionminor{0}
\newcommand\thisgwtcversionfull{\thisgwtcversionmajor.\thisgwtcversionminor}
\newcommand\thisgwtcversion\thisgwtcversionfull
\newcommand\thisgwtc{\gwtc[\thisgwtcversion]}

%% file: gwtc-common-files__macros__symbol_macros.tex
\newcommand{\Msun}{\ensuremath{\mathit{M_\odot}}}

\newcommand{\massone}{\ensuremath{m_1}}
\newcommand{\masstwo}{\ensuremath{m_2}}
\newcommand{\Mc}{\ensuremath{\mathcal{M}}}
\newcommand{\Mtot}{\ensuremath{M}}
\newcommand{\massratio}{\ensuremath{q}}

\newcommand{\chieff}{\ensuremath{\chi_\mathrm{eff}}}

\newcommand{\redshift}{\ensuremath{z}}

\newcommand\PEpdfp{\ensuremath{p}}

\newcommand\PEpdf[2][?]{\ensuremath{\PEpdfp({#2}\ifx#1?\else | {#1}\fi)}}

\newcommand{\SNR}[0]{\ac{SNR}\xspace}

%% file: gwtc-common-files__macros__misc_macros.tex
\newcommand{\soft}[1]{\textsc{#1}}

\newcommand{\GSTLAL}{\soft{GstLAL}\xspace}
\newcommand{\PYCBC}{\soft{PyCBC}\xspace}
\newcommand{\MBTA}{\soft{MBTA}\xspace}
\newcommand{\LALSUITE}{\soft{LALSuite}\xspace}
\newcommand{\NUMPY}{\soft{NumPy}\xspace}
\newcommand{\SCIPY}{\soft{SciPy}\xspace}
\newcommand{\MATPLOTLIB}{\soft{Matplotlib}\xspace}
\newcommand{\SEABORN}{\soft{seaborn}\xspace}
\newcommand{\GWPY}{\soft{GWpy}\xspace}
\newcommand{\GRACEDB}{\soft{GraceDB}\xspace}
\newcommand{\MANIFOLD}{\soft{manifold}\xspace}

%% file: gwtc-common-files__macros__datetime_macros.tex
\newcommand{\OfourBStartDate}{{{2024~April~10}}}
\newcommand{\OfourBStartTime}{{{15:00:00}}}
\newcommand{\OfourBEndDate}{{{2025~January~28}}}
\newcommand{\OfourBEndTime}{{{17:00:00}}}
\newcommand{\OfourEronesixStartDate}{{{2024~March~20}}}
\newcommand{\OfourEronesixStartTime}{{{15:00:00}}}
\newcommand{\OfourEronesixEndDate}{{{2024~April~10}}}
\newcommand{\OfourEronesixEndTime}{{{15:00:00}}}

%% file: gwtc-common-files__macros__quantities_macros.tex
\DeclareSIUnit\parsec{pc}
\DeclareSIUnit\Mpc{\mega\parsec}
\DeclareSIUnit\yr{yr}
\DeclareSIUnit\GpcCubedYear{\giga\parsec\cubed\yr}

%% file: acronyms.tex
\acrodef{LVC}[LVC]{LIGO Scientific and Virgo Collaboration}
\acrodef{aVirgo}{Advanced Virgo}

\acrodef{BBH}[BBH]{binary black hole}
\acrodef{BNS}[BNS]{binary neutron star}
\acrodef{IMBH}[IMBH]{intermediate-mass black hole}
\acrodef{BHNS}[BHNS]{black hole--neutron star binaries}
\acrodef{NSBH}[NSBH]{neutron star--black hole binary}
\acrodef{PBH}[PBH]{primordial black hole}
\acrodef{DBH}[DBH]{dark black hole}
\acrodef{SSM}[SSM]{sub-solar mass}
\acrodef{EB}[EB]{early binaries}
\acrodef{LB}[LB]{late binaries}

\acrodef{MCMC}[MCMC]{Markov chain Monte Carlo}

\newcommand{\GWPOPULATION}{\textsc{GWPopulation}\xspace}

\newcommand{\PANDAS}{\textsc{Pandas}\xspace}

%% file: results_numbers_macros.tex
\newcommand{\GWTCFiveNumEvents}{\reviewed{\ensuremath{390}}\xspace}

\newcommand{\MinMassNSHotMatter}{\reviewed{\ensuremath{0.7~\Msun}\xspace}}
\newcommand{\MinMassNSColdMatter}{\reviewed{\ensuremath{0.1~\Msun}\xspace}}
\newcommand{\MinMassNSObserved}{\reviewed{\ensuremath{1.174\pm 0.004~\Msun}\xspace}}
\newcommand{\FastestBNSName}{\reviewed{PSR J1946+2052}\xspace}
\newcommand{\FastestBNSPeriod}{\reviewed{\ensuremath{16.96~\mathrm{ms}}\xspace}}
\newcommand{\FastestBNSSpinMag}{\reviewed{\ensuremath{0.042}\xspace}}
\newcommand{\FastestSsmBNSSpinMag}{\reviewed{\ensuremath{0.12}\xspace}}
\newcommand{\BrownMomentInertia}{\reviewed{\ensuremath{1.5\times 10^{45}~\mathrm{g}\,\mathrm{cm}^2}\xspace}}

\newcommand{\SignificantOPA}{\reviewed{\ensuremath{2~\mathrm{yr}^{-1}}}\xspace}

\newcommand{\MOneMIN}{\reviewed{0.2~\Msun}\xspace} 
\newcommand{\MOneMAX}{\reviewed{10.0~\Msun}\xspace}
\newcommand{\MTwoMIN}{\reviewed{0.2~\Msun}\xspace}
\newcommand{\MTwoMAX}{\reviewed{1.0~\Msun}\xspace}
\newcommand{\MassRatioMin}{\reviewed{0.1}\xspace} 
\newcommand{\MassRatioMax}{\reviewed{1.0}\xspace}
\newcommand{\MaxSpinLowMass}{\reviewed{0.1}\xspace}
\newcommand{\MaxSpinHighMass}{\reviewed{0.9}\xspace}
\newcommand{\SSMLowHighMassGap}{\reviewed{0.5}\xspace} 
\newcommand{\MinMatch}{\reviewed{0.97}\xspace}
\newcommand{\MinSensitiveFrequency}{\reviewed{15}\xspace}
\newcommand{\SSMFLOW}{\reviewed{\ensuremath{45~\mathrm{Hz}}}\xspace}

\newcommand{\MaxGSTLALOnlineSSMSpin}{\reviewed{0.3}\xspace}
\newcommand{\GSTLALLowMassWaveformApproximant}{\reviewed{TaylorF2}\xspace}
\newcommand{\GstLALTemplateBankWaveformApproximant}{\reviewed{\textsc{IMRPhenomD}}\xspace}
\newcommand{\GSTLALHighMassWaveformApproximant}{\reviewed{SEOBNRv4ROM}\xspace}
\newcommand{\GSTLALSSMMinMatch}{\reviewed{0.965}\xspace}
\newcommand{\GstlalFLOW}{\reviewed{\ensuremath{45~\mathrm{Hz}}}\xspace}
\newcommand{\GstlalFHIGH}{\reviewed{\ensuremath{1024~\mathrm{Hz}}}\xspace}
\newcommand{\GstlalMAXDUR}{\reviewed{\ensuremath{128~\mathrm{s}}}\xspace}
\newcommand{\GstlalBankSize}{\reviewed{\ensuremath{3.5~\mathrm{million}}}\xspace}
\newcommand{\GstlalDtDphiFHIGH}{\reviewed{\ensuremath{1024~\mathrm{Hz}}}\xspace}
\newcommand{\GstlalDtDphiFHIGHOfoura}{\reviewed{\ensuremath{512~\mathrm{Hz}}}\xspace}

\newcommand{\MBTAWaveformApproximant}{\reviewed{TaylorF2}\xspace}
\newcommand{\MBTASEARCHAPPROXLOWMASS}{\reviewed{SpinTaylorT4}\xspace}
\newcommand{\MBTASEARCHAPPROXHIGHMASS}{\reviewed{IMRPhenomTP}\xspace}
\newcommand{\MBTABankSize}{\reviewed{\ensuremath{2.3~\mathrm{million}}}\xspace}
\newcommand{\MBTABankFLOW}{\reviewed{\ensuremath{45~\mathrm{Hz}}}\xspace}
\newcommand{\MBTABankFHIGH}{\reviewed{\ensuremath{1024~\mathrm{Hz}}}\xspace}
\newcommand{\MBTAFilterFLOW}{\reviewed{\ensuremath{35~\mathrm{Hz}}}\xspace}
\newcommand{\MBTAFilterFHIGH}{\reviewed{\ensuremath{512~\mathrm{Hz}}}\xspace}

\newcommand{\PyCBCWaveformApproximant}{\reviewed{TaylorF2}\xspace}
\newcommand{\PyCBCBankSize}{\reviewed{\ensuremath{2.6~\mathrm{million}}}\xspace}
\newcommand{\PyCBCBankFLOW}{\reviewed{\ensuremath{45~\mathrm{Hz}}}\xspace}
\newcommand{\PyCBCSSMMinMatch}{\reviewed{0.975}\xspace}
\newcommand{\PyCBCFilterFLOW}{\reviewed{\ensuremath{30~\mathrm{Hz}}}\xspace}
\newcommand{\PyCBCMaxTempltDur}{\reviewed{\ensuremath{1008~\mathrm{s}}}\xspace}
\newcommand{\PyCBCFilteringWaveformApproximant}{\reviewed{SEOBNRv4ROM}\xspace}
\newcommand{\PyCBCMtotThrforApprox}{\reviewed{4.0~\Msun}\xspace}

\newcommand{\NumberOfPipelinesOOne}{\reviewed{1}\xspace}
\newcommand{\NameOfPipelinesOOne}{\reviewed{\GSTLAL}}
\newcommand{\NumberOfPipelinesOTwo}{\reviewed{1}\xspace}
\newcommand{\NameOfPipelinesOTwo}{\reviewed{\GSTLAL}}
\newcommand{\NumberOfPipelinesOThreeA}{\reviewed{3}\xspace}
\newcommand{\NameOfPipelinesOThreeA}{\reviewed{\GSTLAL, \MBTA, \PYCBC}}
\newcommand{\NumberOfPipelinesOThreeB}{\reviewed{3}\xspace}
\newcommand{\NameOfPipelinesOThreeB}{\reviewed{\GSTLAL, \MBTA, \PYCBC}}
\newcommand{\NumberOfPipelinesOFourA}{\reviewed{3}\xspace}
\newcommand{\NameOfPipelinesOFourA}{\reviewed{\GSTLAL, \MBTA, \PYCBC}}
\newcommand{\NumberOfPipelinesOFourB}{\reviewed{3}\xspace}
\newcommand{\NameOfPipelinesOFourB}{\reviewed{\GSTLAL, \MBTA, \PYCBC}}
\newcommand{\TrialsFactorAcrossObservingRuns}{\reviewed{14}\xspace}

\newcommand{\GSTLALFirstTriggerSEventName}{\reviewed{S240422ed}\xspace}
\newcommand{\GSTLALFirstTriggerPYCBCAllSkyOnlineTriggerFAR}{\reviewed{\ensuremath{10.3~\mathrm{yr^{-1}}}}\xspace}
\newcommand{\GSTLALFirstTriggerGSTLALAllSkyOnlineTriggerFAR}{\reviewed{\ensuremath{{10}^{-5}~\mathrm{yr^{-1}}}}\xspace}
\newcommand{\GSTLALFirstTriggerGSTLALSSMOfflineTriggerFAR}{\reviewed{\ensuremath{0.015~\mathrm{yr^{-1}}}}\xspace}
\newcommand{\GSTLALFirstTriggerGSTLALSSMOfflineTriggerFAP}{\reviewed{0.011}\xspace}
\newcommand{\GSTLALFirstTriggerTime}{\reviewed{2024 April 22 21:35:13.42 UTC}\xspace}
\newcommand{\GSTLALFirstTriggerOnlineMchirp}{\reviewed{$\sim$ 1.7} $M_\odot$\xspace}
\newcommand{\GSTLALFirstTriggerOnlineChieff}{\reviewed{$\gtrsim$ 0.75}\xspace}
\newcommand{\GSTLALFirstTriggerGSTLALAllSkyOnlineTriggerFARUpdated}{\reviewed{$\sim$ 1 per 35 days}\xspace}

\newcommand{\MBTAFirstTriggerSEventName}{\reviewed{S241229bq}\xspace}
\newcommand{\MBTAFirstTriggerTime}{\reviewed{2024 December 29 13:52:36.83 UTC}\xspace}
\newcommand{\MBTAFirstTriggerFAR}{\reviewed{\ensuremath{0.0043~\mathrm{yr^{-1}}}}\xspace}

\newcommand{\MBTAFirstTriggerMBTASSMOnlineTriggerFAR}{\reviewed{\ensuremath{570~\mathrm{yr^{-1}}}}\xspace}
\newcommand{\MBTAFirstTriggerGSTLALSSMOnlineTriggerFAR}{\reviewed{\ensuremath{100~\mathrm{yr^{-1}}}}\xspace}
\newcommand{\MBTAFirstTriggerGSTLALSSMOfflineTriggerFAR}{\reviewed{\ensuremath{34~\mathrm{yr^{-1}}}}\xspace}
\newcommand{\MBTAFirstTriggerPyCBCSSMOfflineTriggerFAR}{\reviewed{\ensuremath{7.7~\mathrm{yr^{-1}}}}\xspace}

\newcommand{\MBTAFirstTriggerTestFlow}{\reviewed{25} Hz\xspace}
\newcommand{\MBTAFirstTriggerFAP}{\reviewed{0.0027}\xspace}
\newcommand{\MBTAFirstTriggerSignificance}{\reviewed{2.8} standard deviations\xspace}
\newcommand{\MBTAFirstTriggerFARdiscrepancy}{\reviewed{$10^{-3}$}\xspace}

\newcommand{\MBTAFirstTriggerFARDiscrepancyCutHigh}{\reviewed{$1~\mathrm{yr}^{-1}$}\xspace}
\newcommand{\MBTAFirstTriggerFARDiscrepancyCutLow}{\reviewed{$0.005~\mathrm{yr}^{-1}$}\xspace}

\newcommand{\GSTLALSNRThreshold}{\reviewed{$4.0$}\xspace}
\newcommand{\MBTASNRThreshold}{\reviewed{$4.4$}\xspace}

\newcommand{\GSTLALTableRow}[9]{%
  \reviewed{\ensuremath{#1}} &
  \GSTLAL &                   
  \reviewed{#2} &             
  \reviewed{\ensuremath{#3}} &
  \reviewed{\ensuremath{#4}} &
  \reviewed{\ensuremath{#5}} &
  \reviewed{\ensuremath{#6}} &
  \reviewed{#7} &             
  \reviewed{#8} &             
  \reviewed{\ensuremath{#9}} \\
}

\newcommand{\GSTLALTrigOneData}{%
  {0.015}        
  {2024 April 22 21:35:13.42 UTC}
  {5.28}           
  {0.85}           
  {0.79}           
  {0.79}           
  {HLV}           
  {HL}            
  {9.74}          
}

\newcommand{\MBTATableRow}[9]{%
  \reviewed{\ensuremath{#1}} &
  \MBTA &                     
  \reviewed{#2} &             
  \reviewed{\ensuremath{#3}} &
  \reviewed{\ensuremath{#4}} &
  \reviewed{\ensuremath{#5}} &
  \reviewed{\ensuremath{#6}} &
  \reviewed{#7} &             
  \reviewed{#8} &             
  \reviewed{\ensuremath{#9}} \\
}

\newcommand{\MBTATrigOneData}{%
  {0.0043}             
  {2024 December 29 13:52:36.83 UTC}
  {1.04}           
  {0.49}           
  {-0.33}       
  {0.040}       
  {HLV}            
  {HL}             
  {10.2}          
}

\newcommand{\MBTATrigTwoData}{%
  {1.3}             
  {2025 January 2 09:44:02.30 UTC}
  {0.32}           
  {0.23}           
  {0.02}           
  {-0.02}          
  {HLV}            
  {LV}             
  {10.6}           
}

\newcommand{\BNSVTorderofmagnitude}{\reviewed{\qty{\sim e-3}{\giga\parsec\cubed\yr}}\xspace}

\newcommand{\GSTLALLoudestTriggerFAR}{\reviewed{\ensuremath{0.015~\mathrm{yr^{-1}}}}\xspace}
\newcommand{\MBTALoudestTriggerFAR}{\reviewed{\ensuremath{0.0043~\mathrm{yr^{-1}}}}\xspace}
\newcommand{\PYCBCLoudestTriggerFARCombined}{\reviewed{\ensuremath{1.3\,\mathrm{yr^{-1}}}}\xspace}

\newcommand{\FourthBObsAnalysisTime}{\reviewed{\ensuremath{0.84\,\mathrm{yr}}}\xspace}
\newcommand{\FourthAObsAnalysisTime}{\reviewed{\ensuremath{0.76\,\mathrm{yr}}}\xspace}

\newcommand{\TotalInjections}{\reviewed{\ensuremath{3.7\times 10^7}\xspace}}
\newcommand{\InjSpacing}{\reviewed{\ensuremath{32\,\mathrm{s}}\xspace}}
\newcommand{\MBTAInjSpacing}{\reviewed{\ensuremath{16\,\mathrm{s}}\xspace}}

\newcommand{\InjMassPowerLawAlpha}{\reviewed{\ensuremath{-1}}\xspace}
\newcommand{\MinPrimaryMassBH}{\reviewed{\ensuremath{0.19~\Msun}\xspace}}
\newcommand{\MaxPrimaryMassBH}{\reviewed{\ensuremath{11.1~\Msun}\xspace}}
\newcommand{\MinSecondaryMassBH}{\reviewed{\ensuremath{0.19~\Msun}\xspace}}
\newcommand{\MaxSecondaryMassBH}{\reviewed{\ensuremath{1.1~\Msun}\xspace}}
\newcommand{\MinPrimaryMassNS}{\reviewed{\ensuremath{0.5~\Msun}\xspace}}
\newcommand{\MaxPrimaryMassNS}{\reviewed{\ensuremath{2.5~\Msun}\xspace}}
\newcommand{\MaxPrimaryMassNSOfoura}{\reviewed{\ensuremath{3.0~\Msun}\xspace}}
\newcommand{\EOSMaxMassNS}{\reviewed{\ensuremath{\sim2.4~\Msun}\xspace}}
\newcommand{\MinSecondaryMassNS}{\reviewed{\ensuremath{0.5~\Msun}\xspace}}
\newcommand{\MaxSecondaryMassNS}{\reviewed{\ensuremath{1.1~\Msun}\xspace}}

\newcommand{\DistributionWeight}{\reviewed{\ensuremath{0.5}}\xspace}
\newcommand{\NormalDistMu}{\reviewed{\ensuremath{0}}\xspace}
\newcommand{\NormalDistSigma}{\reviewed{\ensuremath{0.1}}\xspace}
\newcommand{\BetaDistAlpha}{\reviewed{\ensuremath{1}}\xspace}
\newcommand{\BetaDistBeta}{\reviewed{\ensuremath{4}}\xspace}
\newcommand{\MinSpinMag}{\reviewed{\ensuremath{0}}\xspace}
\newcommand{\MaxSpinMag}{\reviewed{\ensuremath{1}}\xspace}

\newcommand{\IMRPHENOMXPHM}{\reviewed{\textsc{IMRPhenomXPHM}}\xspace}
\newcommand{\IMRPHENOMXPNRTIDALVTHREE}{\reviewed{\textsc{IMRPhenomXP\_NRTidalv3}}\xspace}
\newcommand{\GPPVADDtwo}{\reviewed{\textsc{GPPVA\_DD2}}\xspace}

\newcommand{\MaxRedshiftInjBH}{\reviewed{\ensuremath{0.25}\xspace}}
\newcommand{\MaxRedshiftInjNS}{\reviewed{\ensuremath{0.2}\xspace}}
\newcommand{\MaxRedshiftInjOfourb}{\reviewed{\ensuremath{0.25}\xspace}}

\newcommand{\HighSpinPBH}{\reviewed{\ensuremath{0.7}}\xspace}

\newcommand{\TotalInjVTbbh}{\reviewed{\ensuremath{3.34\,\mathrm{Gpc^{3}\,yr}}}\xspace}
\newcommand{\TotalInjVTbns}{\reviewed{\ensuremath{1.83\,\mathrm{Gpc^{3}\,yr}}}\xspace}

\newcommand{\FourthAGSTLALLoudestTriggerFAR}{\reviewed{\ensuremath{1.2\,\mathrm{yr^{-1}}}}}
\newcommand{\FourthAMBTALoudestTriggerFAR}{\reviewed{\ensuremath{0.87\,\mathrm{yr^{-1}}}}}
\newcommand{\FourthAPYCBCLoudestTriggerFAR}{\reviewed{\ensuremath{1.3\,\mathrm{yr^{-1}}}}}

\newcommand{\MinMcVT}{\reviewed{0.16~\Msun}\xspace}
\newcommand{\MaxMcVT}{\reviewed{2.72~\Msun}\xspace}
\newcommand{\MaxLowSpinMagVTpop}{\reviewed{\ensuremath{0.1}}\xspace}
\newcommand{\MaxHighSpinMagVTpop}{\reviewed{\ensuremath{0.7}}\xspace}
\newcommand{\MinMassRatioVTpop}{\reviewed{\ensuremath{0.05}}\xspace}
\newcommand{\MinMOneVT}{\reviewed{0.19~\Msun}\xspace}
\newcommand{\MaxMOneVT}{\reviewed{11.1~\Msun}\xspace}
\newcommand{\MinMTwoVT}{\reviewed{0.19~\Msun}\xspace}
\newcommand{\MaxMTwoVT}{\reviewed{1.1~\Msun}\xspace}

\newcommand{\missedInjectionGstLALPyCBC}{\reviewed{\ensuremath{1.1\%}}\xspace}
\newcommand{\missedInjectionMBTA}{\reviewed{\ensuremath{0.7\%}}\xspace}
\newcommand{\maxIncreaseAtoBGstLALPyCBC}{\reviewed{33\%}\xspace}
\newcommand{\maxIncreaseAtoBMBTA}{\reviewed{73\%}\xspace}
\newcommand{\TobsIncrease}{\reviewed{11\%}\xspace}
\newcommand{\EfficiencyIncrease}{\reviewed{50\%}\xspace}
\newcommand{\increaseTobsABtoA}{\reviewed{2.1}\xspace}
\newcommand{\increaseVtABtoA}{\reviewed{3}\xspace}
\newcommand{\ratioHightoLowSpinsRunB}{\reviewed{25\%}\xspace}
\newcommand{\ratioHightoLowSpinsRunAB}{\reviewed{20\%}\xspace}
\newcommand{\increaseVarianceHighSpins}{\reviewed{6}\xspace}

\newcommand{\lastMcbinmin}{\reviewed{2.44}}
\newcommand{\lastMcbinmax}{\reviewed{2.72}}

\newcommand{\highvariancemOnemin}{\reviewed{6.74}}
\newcommand{\highvariancemOnemax}{\reviewed{8.92}}
\newcommand{\highvariancemTwomin}{\reviewed{0.19}}
\newcommand{\highvariancemTwomax}{\reviewed{0.37}}

\newcommand{\BBHVTFARThreshold}{\reviewed{\ensuremath{1~{\rm yr^{-1}}}}\xspace}
\newcommand{\BNSVTFARThreshold}{\reviewed{\ensuremath{1~{\rm yr^{-1}}}}\xspace}

\newcommand{\MaxSecondaryMassNSVT}{\reviewed{\ensuremath{1.0~\Msun}\xspace}}

\newcommand{\increaseLowBoundBNSvt}{\reviewed{\ensuremath{2.2}\xspace}}
\newcommand{\increaseHighBoundBNSvt}{\reviewed{\ensuremath{3}\xspace}}
\newcommand{\lowUncertBNSVT}{\reviewed{\ensuremath{1.2\%}\xspace}}
\newcommand{\highUncertBNSVT}{\reviewed{\ensuremath{5.2\%}\xspace}}
\newcommand{\fractionBBHtoBNS}{\reviewed{\ensuremath{0.8}\xspace}}

\newcommand{\LogNormSigmaBNSVT}{\reviewed{\ensuremath{0.1~\Msun}\xspace}}
\newcommand{\maxspinBNS}{\reviewed{\ensuremath{0.4}}\xspace}

\newcommand{\RateULMin}{\reviewed{\ensuremath{50\,{\rm Gpc^{-3}\,yr^{-1}}}}\xspace}
\newcommand{\RateULMax}{\reviewed{\ensuremath{5000\,{\rm Gpc^{-3}\,yr^{-1}}}}\xspace}

\newcommand{\MostConstrainedRateMcGSTLAL}{\reviewed{\ensuremath{66.8\,{\rm Gpc^{-3}\,yr^{-1}}}}}
\newcommand{\MostConstrainedRateMcMBTA}{\reviewed{\ensuremath{72.0\,{\rm Gpc^{-3}\,yr^{-1}}}}}
\newcommand{\MostConstrainedRateMcPYCBC}{\reviewed{\ensuremath{50.0\,{\rm Gpc^{-3}\,yr^{-1}}}}}

\newcommand{\MostConstrainedRateCompmassGSTLAL}{\reviewed{\ensuremath{65.6\,{\rm Gpc^{-3}\,yr^{-1}}}}\xspace}
\newcommand{\MostConstrainedRateCompmassMBTA}{\reviewed{\ensuremath{68.0\,{\rm Gpc^{-3}\,yr^{-1}}}}\xspace}
\newcommand{\MostConstrainedRateCompmassPYCBC}{\reviewed{\ensuremath{47.8\,{\rm Gpc^{-3}\,yr^{-1}}}}\xspace}

\newcommand{\SSMBNSRateOFourAB}{\reviewed{\ensuremath{249\,{\rm Gpc^{-3}\,yr^{-1}}}}\xspace}
\newcommand{\GWTCfourBNSRate}{\reviewed{\ensuremath{(8.8\text{--}250)\,{\rm Gpc^{-3}\,yr^{-1}}}}\xspace}

\newcommand{\LBfPBHBelowOneOfourAB}{\reviewed{\ensuremath{(0.62\text{--}0.74)\,\Msun}}\xspace}
\newcommand{\MaxMPBH}{\reviewed{1\,\Msun}\xspace}
\newcommand{\MinMPBH}{\reviewed{0.35\,\Msun}\xspace}
\newcommand{\EBfPBHMaxMPBHOfourAB}{\reviewed{\ensuremath{(0.018\text{--}0.032)}}\xspace}
\newcommand{\EBfPBHMinMPBHOfourAB}{\reviewed{\ensuremath{(0.10\text{--}0.18)}}\xspace}

\newcommand{\FARThreshVTDBH}{\reviewed{\ensuremath{0.01~\mathrm{yr^{-1}}}}\xspace}
\newcommand{\DBHConstraintOfouraBest}{\reviewed{\ensuremath{(1.8\text{--}2.5) \times 10^{-5}}}\xspace}
\newcommand{\DBHConstraintOfourbBest}{\reviewed{\ensuremath{(1.0\text{--}1.1) \times 10^{-5}}}\xspace}
\newcommand{\DBHConstraintBestMminOfoura}{\reviewed{\ensuremath{(0.65\text{--}0.80)\,\Msun}}\xspace}
\newcommand{\DBHConstraintBestMminOfourb}{\reviewed{\ensuremath{(0.65\text{--}0.85)\,\Msun}}\xspace}
\newcommand{\DBHConstraintFactorOfouraToOfourb}{\reviewed{\ensuremath{2}\xspace}}
\newcommand{\DBHMminOfouraBound}{\reviewed{\ensuremath{(4.2\text{--}4.8) \times 10^{-2}\,\Msun}}\xspace}
\newcommand{\DBHMminOfourbBound}{\reviewed{\ensuremath{(4.2\text{--}4.4) \times 10^{-2}\,\Msun}}\xspace}
\newcommand{\darkprotonlower}{\reviewed{2.4}\xspace}
\newcommand{\darkprotonOfouraupper}{\reviewed{\ensuremath{(10.8\text{--}11.5)}}\xspace}
\newcommand{\darkprotonOfourbupper}{\reviewed{\ensuremath{(11.3\text{--}11.7)}}\xspace}
\newcommand{\DBHMcBinCount}{\reviewed{\ensuremath{99}}\xspace} 
\newcommand{\DBHMcBinLo}{\reviewed{\ensuremath{0.16\,\Msun}}\xspace}
\newcommand{\DBHMcBinHi}{\reviewed{\ensuremath{2.72\,\Msun}}\xspace}

\newcommand{\darkprotonupperold}{\reviewed{\ensuremath{16.5}\text{--}\ensuremath{16.6}}}
\newcommand{\mminlowerboundold}{\reviewed{\ensuremath{0.020\,\Msun}\text{--}\ensuremath{0.021\,\Msun}}\xspace}
\newcommand{\fconstraintOfouraold}{\reviewed{(0.9\,\mathrm{to}\,1.2) \times 10^{-5}}}

\newcommand{\DBHtmerger}{\reviewed{\ensuremath{10\,\mathrm{Gyr}}}\xspace}
\newcommand{\DBHfbinary}{\reviewed{\ensuremath{0.26}}\xspace}

\newcommand{\DBHrhoDM}{\reviewed{\ensuremath{3.3 \times 10^{19}\,\Msun~\mathrm{Gpc}^{-3}}}\xspace}

%% file: authorlist__LSC-Virgo-KAGRA-Authors-Feb-2026-aas-fragment.tex
\author{A.~G.~Abac\,\orcidlink{0000-0003-4786-2698}}
\affiliation{Max Planck Institute for Gravitational Physics (Albert Einstein Institute), D-14476 Potsdam, Germany}
\author{A.~Abe}
\affiliation{Department of Physics, Graduate School of Science, Osaka Metropolitan University, 3-3-138 Sugimoto-cho, Sumiyoshi-ku, Osaka City, Osaka 558-8585, Japan  }
\author{I.~Abouelfettouh}
\affiliation{LIGO Hanford Observatory, Richland, WA 99352, USA}
\author{F.~Acernese}
\affiliation{Dipartimento di Fisica ``E.R. Caianiello'', Universit\`a di Salerno, I-84084 Fisciano, Salerno, Italy}
\affiliation{INFN, Sezione di Napoli, I-80126 Napoli, Italy}
\author{K.~Ackley\,\orcidlink{0000-0002-8648-0767}}
\affiliation{University of Warwick, Coventry CV4 7AL, United Kingdom}
\author{A.~Adam}
\affiliation{OzGrav, University of Western Australia, Crawley, Western Australia 6009, Australia}
\author{S.~Adhicary\,\orcidlink{0009-0004-2101-5428}}
\affiliation{The Pennsylvania State University, University Park, PA 16802, USA}
\author{D.~Adhikari}
\affiliation{Max Planck Institute for Gravitational Physics (Albert Einstein Institute), D-30167 Hannover, Germany}
\affiliation{Leibniz Universit\"{a}t Hannover, D-30167 Hannover, Germany}
\author{R.~X.~Adhikari\,\orcidlink{0000-0002-5731-5076}}
\affiliation{LIGO Laboratory, California Institute of Technology, Pasadena, CA 91125, USA}
\author{V.~K.~Adkins}
\affiliation{Louisiana State University, Baton Rouge, LA 70803, USA}
% ...

%% file: source__abstract.tex
We report on a gravitational wave search for compact binary coalescences 
with secondary mass between $0.2\,\Msun$ and $1\,\Msun$ and primary mass between
$0.2\,\Msun$ and $10\,\Msun$, restricting the component mass ratio to lie between 0.1 and 1. 
The analysis uses data collected by the LIGO and Virgo detectors between 
\OfourBStartDate\ \OfourBStartTime\ and \OfourBEndDate\ \OfourBEndTime, 
the second part of the fourth observing run of the LIGO--Virgo--KAGRA Collaboration.
Results from two out of three search pipelines show outlier candidates, but not 
at a sufficient level of significance to provide evidence for detection.
We report the combined sensitivity of the current searches over the first and second 
parts of the observing run to the target sub-solar mass black hole population,
as well as to neutron star binaries with one or more sub-solar component.
We place upper limits on the merger rate of sub-solar mass black holes, ranging from
\RateULMin~to~\RateULMax ($90\%$ confidence) 
for different component mass combinations.
We use the merger rate limits to constrain two illustrative dark matter scenarios 
that can form sub-solar mass compact objects: primordial black holes, and dark 
black holes forming in a dissipative dark matter model.
Our search limits the fraction of dark matter in primordial black holes, assuming a
monochromatic mass distribution, to be $\leq$ \EBfPBHMaxMPBHOfourAB at \MaxMPBH,
and $\leq$ \EBfPBHMinMPBHOfourAB at \MinMPBH for early-forming binaries; or 
to be less than $1$ at masses above \LBfPBHBelowOneOfourAB for late-forming binaries.
For the dissipative model, we constrain the fraction of dark matter for a range of
minimum dark black hole masses below $1\,\Msun$ to be as low as \DBHConstraintOfourbBest;
we also exclude minimum masses above \DBHMminOfourbBound, if dark black holes constitute
all of the dark matter.
For binary neutron stars that include sub-solar mass components, assuming a simple,
fixed population model we find a merger rate upper limit of \SSMBNSRateOFourAB.

%% file: source__1-introduction.tex
\section{Introduction}\label{sec:1-introduction}

The \acl{LVK}~\citep[\acsu{LVK};][]{KAGRA:2013rdx} has published over  
\GWTCFiveNumEvents 
probable \ac{GW} candidates from searches for \acp{CBC} in data from the 
\acl{O1}~\citep[\acsu{O1};][]{2019PhRvX...9c1040A}, 
\acl{O2}~\citep[\acsu{O2};][]{2021PhRvX..11b1053A,2024PhRvD.109b2001A}, 
\acl{O3}~\citep[\acsu{O3};][]{2023PhRvX..13d1039A}, the 
\acl{O4a}~\citep[\acsu{O4a};][]{2026ApJ..1004L..22A,2026ApJ..1004..232A}, 
and the \acl{O4b}~\citep[\acsu{O4b};][]{2026arXiv260527223T, 2026arXiv260527224T, 2026arXiv260527225T}
of the 
\acl{LIGO}~\citep[\acsu{LIGO};][]{2025PhRvD.111f2002C,2023PhRvX..13d1021G,2024Sci...385.1318J,2025CQGra..42h5016S,2016PhRvD..93k2004M,Harry:2010zz,2016PhRvL.116m1103A} 
and Virgo~\citep{2015CQGra..32b4001A, 2026arXiv260719273V} detectors in the latest \ac{GWTC} -- 
GWTC-5.0~\citep{2026arXiv260527225T}. 
These candidates are produced by searches targeting the inspiral and merger of 
binary systems of compact object remnants of massive stars. 
Namely, \acp{BBH}, \acp{BNS} and \acp{NSBH} with component masses $\geq 1\,\Msun$, 
where the lower bound is motivated observationally by known 
pulsars~\citep{2015ApJ...812..143M, Martinez:2017jbp, Lazarus:2016hfu, 2018ApJ...854L..22S}, 
and theoretically by the 
expectation that \ac{NS} masses should lie near the Chandrasekhar limit for 
white dwarfs of $1.4\,\Msun$ as determined by the nucleon masses~\citep{Chandrasekhar:1931ih}.
Some neutron stars observed in X-ray binaries have masses of 
$\sim 1\,\Msun$~\citep{vanderMeer:2007hj,Lattimer:2004sa}.  
Additionally, \acp{BH} produced by the gravitational collapse of massive stars
are expected to have masses exceeding those of \ac{NS}.

Although \ac{GW} detectors are most sensitive to mergers of stellar-mass 
compact objects, they can also detect signals from less massive binaries of 
compact objects, if such systems exist in nature~\citep{Magee:2018opb}.
Targeted searches for \ac{SSM} compact objects open a window into exotic 
formation channels that may involve physics beyond the standard 
astrophysical paradigm~\citep{Singh:2024fdd}.
A key motivation for these searches is the possibility that some compact 
objects could be \acp{PBH}, which are a candidate for dark 
matter~\citep{Hawking:1971ei,Chapline:1975ojl,Green:2020jor}.
\acp{PBH} have been proposed to contribute to the population of \ac{GW} sources 
since the first detection of \acp{GW}: 
GW150914~\citep{2016PhRvL.116f1102A,Bird:2016dcv,Sasaki:2016jop,Clesse:2016vqa}.
They are purported to form in the early universe, and to dominate the merger rate 
across all redshifts with a peak around \redshift $\geq 30$, compared to the 
merger rate of stellar \acp{BH}, which, if it follows the star formation rate, 
would peak around \redshift $\sim$ 1--2~\citep{Ng:2022agi,Ng:2022vbz}.
Although \acp{PBH} can populate a wide mass spectrum extending above a solar 
mass~\citep{Carr:2019kxo}, we rely on the detection of a \ac{SSM} candidate to conclusively 
distinguish them from stellar \acp{BH}.

\Ac{GW} observations of \acp{BBH} can also constrain theories of dissipative 
dark matter and its chemistry, where dark halos collapse under gravity to form 
\acp{DBH}~in regions of sufficiently high dark matter 
density~\citep{Shandera:2018xkn,2023MNRAS.524.5984L,Singh:2020wiq}.
The merger rate of \acp{DBH} in this scenario is inversely proportional to the 
average~\ac{DBH} mass.
Therefore, \ac{SSM} \acp{BH} forming from dissipative halos could have a higher 
predicted merger rate than \acp{BH} forming through regular stellar 
evolutionary channels, which implies that \ac{SSM} searches can provide strict 
limits on dissipative dark matter models~\citep{Shandera:2018xkn}.
Especially if the lowest possible \ac{DBH} mass is smaller than the white dwarf 
Chandrasekhar limit, we can start to probe the particle nature of dissipative 
dark matter using \ac{SSM} searches.

Additionally, models where accumulation of non-annihilating dark matter 
transmutes \acp{NS} to \acp{BH} with masses below the maximum 
\ac{NS} mass have been constrained using limits on merger rates from  
\ac{SSM} searches.
Non-detections of these low-mass \acp{BH} provide upper limits on the 
interaction cross-section of dark matter particles with nucleons for GeV scale 
dark matter~\citep{Dasgupta:2020mqg,Bhattacharya:2023stq}.

Searches for binaries with \ac{SSM} compact objects have been performed in
data from \ac{LIGO} and Virgo since Initial LIGO's science runs \citep{LIGOScientific:2005fbz, 
LIGOScientific:2007npa, 2018PhRvL.121w1103A, 2019PhRvL.123p1102A, 
2022PhRvL.129f1104A, 2023MNRAS.524.5984L, LIGOScientific:2026wxz, 2021ApJ...915...54N, 2021PhRvL.126b1103N, Phukon:2021cus, 2022PhRvD.106b3024N, Kacanja:2024hme, Kacanja:2026byy}.
\ac{LVK} matched-filtering \ac{SSM} searches have probed masses down to $0.2\,\Msun$; this 
parameter space was initially motivated by gravitational microlensing surveys 
of the Large Magellanic Cloud which suggested that $\sim 20\%$ of the Galactic 
halo is composed of compact objects with masses in the range $0.15\,\Msun$ to 
$0.9\,\Msun$~\citep{MACHO:2000qbb}.
Since \ac{O2}, our \ac{SSM} searches have allowed for spinning components, 
initially with dimensionless spin magnitudes $\leq 
0.1$~\citep{2019PhRvL.123p1102A}.
In \ac{O3}, we allowed the primary mass to extend up to $10\,\Msun$ in order to
include systems with more asymmetric masses, while also increasing the maximum 
dimensionless spin magnitude to $0.9$ for compact objects with component masses 
greater than $0.5\,\Msun$~\citep{2022PhRvL.129f1104A, 2023MNRAS.524.5984L}.
We keep these configuration choices unchanged in our search for \ac{SSM} candidates
in \ac{O4a} and \ac{O4b} data.  
Although \ac{PBH} and \ac{DBH} formation scenarios can populate the mass 
spectrum below $0.2\,\Msun$, the target parameter space is restricted to 
alleviate the computational cost of \ac{CBC} matched-filtering searches, and given 
that the horizon distance of \ac{LIGO} and Virgo detectors decreases with 
decreasing mass~\citep{2018PhRvL.121w1103A, 2012PhRvD..85l2006A, Fairhurst:2007qj}.

Searches for \acp{SSM} have been conducted on publicly available data with 
parameter spaces that extend outside what is explored by the \ac{LVK} searches. 
No detection has been made to date by any of the searches.
With \ac{O1} and \ac{O2} data, \citet{2021ApJ...915...54N} reports a search 
for \ac{SSM} mergers with nonnegligible orbital eccentricity, for component 
masses spanning $0.1-7\,\Msun$ (primary) and $0.1-1\,\Msun$ (secondary), and 
\citet{2021PhRvL.126b1103N} reports a search for mergers of extreme mass-ratio
binaries for component mass ranges of $20-100\,\Msun$ and $0.01-1\,\Msun$. 
\citet{Phukon:2021cus} reports a search for mergers of a sub-solar mass component with a 
component with mass greater than $2\,\Msun$ in \ac{O2}.
\citet{2022PhRvD.106b3024N} reports searches on \ac{O3} data where the secondary 
component is sub-solar mass, while the primary mass extends from $0.1-100\,\Msun$, 
and orbital eccentricity is allowed for limited mass regions. 
Using \ac{O3} data,~\citet{Kacanja:2024hme} reports a targeted search for sub-solar 
mass \ac{NS} binaries with component masses of $0.1-2\,\Msun$ and $0.1-1\,\Msun$, 
allowing for nonzero tidal deformabilities. 
Finally,~\citet{Kacanja:2026byy} reports the search for \ac{SSM} mergers in \ac{O4a} 
data, where the component masses are $0.1-2\,\Msun$ and $0.1-1\,\Msun$ and 
tidal deformabilities are included for low mass \acp{NS}. 

Searches for planetary-mass compact object binaries have also been 
conducted recently over \ac{O3} and \ac{O4a} data, taking advantage of the quasi-monochromatic 
evolution of the emitted \ac{GW} in the \ac{LVK} bandwidth to set limits on the
abundance of $[10^{-7},10^{-2}]\,\Msun$ \acp{PBH}~\citep{Miller:2024fpo, LIGOScientific:2025vwc};
in addition, searches for continuous gravitational waves in \ac{O4} data placed
limits on \ac{PBH} formation considering
\ac{SSM} \acp{BH} in the asteroid-mass region~\citep{LIGOScientific:2026plm}.
The absence of detections in previous runs has provided useful constraints 
on the abundance of \acp{PBH} and on dissipative dark matter 
models~\citep{2023MNRAS.524.5984L, LIGOScientific:2026wxz}.

Low-latency \ac{SSM} searches were operated for the first time in \ac{O4a}, 
motivated by the searches' sensitivity to typical electromagnetically-bright sources.
These real-time searches were continued throughout \ac{O4} in preparation for
enabling prompt follow-up of significant candidates~\citep{2025CQGra..42j5009A,Hanna:2024tom}.
However, public alerts from low-latency \ac{SSM} searches were first issued
in \ac{O4c}, thus the corresponding candidates lie out of the scope of the
present study.

By contrast, archival (offline) searches can determine the significance of
candidates using the noise background collected over an entire observing run, and 
can undertake large-scale simulation analyses for sensitivity estimation. 
In this paper, we present results from the archival \ac{SSM} searches 
performed by three pipelines: \GSTLAL, \MBTA and \PYCBC. 
We report their detection sensitivities for \ac{SSM} \acp{BBH} and \acp{BNS} by 
repeating the searches on strain data from \ac{O4b} with simulated signals added.
By combining the search sensitivities to \ac{SSM} signals in \ac{O4a} and \ac{O4b}, 
we derive limits on the merger rate of \ac{SSM} \acp{BBH}, which are 
used to update the constraints on the fraction of dark matter in \acp{PBH} and 
the fraction of dark matter in \acp{DBH}.
For the \ac{PBH} scenario, we present constraints across the component-mass 
plane in addition to previously used point-mass 
distributions~\citep{2018PhRvL.121w1103A, 2019PhRvL.123p1102A, 
2022PhRvL.129f1104A, 2023MNRAS.524.5984L, LIGOScientific:2026wxz}.

In addition to our constraints on specific dark matter models, following
\citet{LIGOScientific:2026wxz} we report the sensitivity of the \ac{SSM}
search to binaries comprising one or two low-mass \acp{NS}.
\ac{SSM} search templates have non-trivial overlap with \ac{BNS} signals from an
astrophysical population, while extending 
into the high-spin parameter space typically excluded in template-bank based 
searches for \ac{BNS}~\citep{2026ApJ..1004L..22A,2026arXiv260527224T}.
Matter effects involving tidal interactions of the binary components influence
the signal phasing starting at an effective fifth post-Newtonian order~\citep{2008PhRvD..77b1502F};
for \acp{BNS} with super-solar masses, tidal deformations affect waveforms at
$\geq$ 1\,kHz, where \ac{LIGO} and Virgo have reduced sensitivity.
However, extremely large tides at low \ac{NS} masses ($\leq 0.5\,\Msun$) cause dephasing below 1\,kHz, which can result in significant mismatches between the \ac{BBH} template waveforms and \ac{BNS} signals~\citep{Bandopadhyay:2022tbi}.
Therefore, we extend our sensitivity estimation to include simulated signals 
from \ac{SSM}~\ac{NS} mergers~\citep{Metzger:2024ujc,Chen:2025uwd} to determine 
the detection efficiency for this population.

\Acp{NS} are formed in the last phases of stellar evolution, due to gravitational 
collapse at the endpoint of nuclear burning~\citep{Burrows:2020qrp}, or in 
binary systems where the accretion to a white dwarf induces collapse~\citep{Nomoto:1990xk}.
Simulations of \ac{NS} 
formation~\citep{2018MNRAS.481.3305S, 2020ApJ...890...51E, Woosley:2020mze} 
predict a resulting minimum \ac{NS} mass higher than $1~\Msun$, comparable to 
the Chandrasekhar limit \citep{Chandrasekhar:1931ih}.
However, lower bounds on possible \ac{NS} masses based on \ac{EOS} analysis 
cover a wide range, largely dependent on the type of \ac{EOS} assumed; bounds 
for dense hot matter models are $\sim \MinMassNSHotMatter$~\citep{Burgio:2010ek}, 
closer to the Chandrasekhar limit, as compared to $\sim \MinMassNSColdMatter$ 
for cold matter models~\citep{Potekhin:2013qqa}.
In the case of spinning \acp{NS}, the rotation would require a higher 
gravitational binding force, hence a higher minimum mass is required to keep
the \ac{NS} stable~\citep{Colpi:1990fe}.
In summary, although the lowest mass \ac{NS} observed to date is 
$\MinMassNSObserved$ ($1\,\sigma$ uncertainty) \citep{2015ApJ...812..143M},
theoretical scenarios for sub-solar \acp{NS} remain open.

Various astrophysical mechanisms have been proposed for formation of \ac{SSM} 
\acp{BNS}: as proposed by~\citet{Clark:1977}, if the less massive \ac{NS} 
reaches the Roche radius, then tidal disruption can be so extreme as to cause 
mass stripping.
As a consequence of mass transfer to the primary, the less massive \ac{NS} 
could therefore fall into the \ac{SSM} regime.
Alternatively, the Chandrasekhar limit, originally derived for the conditions 
found in stellar cores, might allow lower \ac{NS} masses if the environment 
is sufficiently neutron-rich.
This situation can be found in unstable disks around a central \ac{BH}, which 
result from the gravitational collapse of rapidly rotating stars~\citep{Piro:2006ja}.
In such disks, gravitational instabilities may lead to the formation of 
sub-solar \acp{NS} from sufficiently neutron-rich 
regions~\citep{Metzger:2024ujc, Chen:2025uwd}.

The structure and analysis framework of this paper closely follow 
the search for binaries with at least one \ac{SSM} compact object in data from 
\ac{O4a}~\citep{LIGOScientific:2026wxz}. 
Here we update those results analyzing the data collected during \ac{O4b}; this 
paper thus presents the most comprehensive results to date of the \ac{LVK}'s 
\ac{SSM} search.

%% file: source__2-search.tex
\section{Searches}\label{sec:2-search}

We report results from three independent search pipelines 
- \GSTLAL, \MBTA, and \PYCBC — that analyzed data collected during \ac{O4b}, 
from \textcolor{black}{\OfourBStartDate{} \OfourBStartTime{}}
to \textcolor{black}{\OfourBEndDate{} \OfourBEndTime{}}, and the preceding engineering run, 
specifically the period with coincident operation of the two \ac{LIGO} 
detectors and Virgo detector from \OfourEronesixStartDate{} \OfourEronesixStartTime{} to 
\OfourEronesixEndDate{} \OfourEronesixEndTime{}~\citep{2026arXiv260527223T}.
While \ac{O4a} and the engineering run preceding \ac{O4a} contained data only 
from the two \ac{LIGO} detectors, the data analyzed for this paper and 
\gwtcfive included data from the Virgo detector, which resumed its observations 
prior to \ac{O4b}. 
Details regarding the sensitivity of detectors during \ac{O4b} can be found 
in \citet{2026arXiv260527223T}, while details of detector characterization and data 
quality in \ac{O4b} are described in \citet{O4bDetchar} and \citet{2026arXiv260527224T}, 
respectively.

All three pipelines employ matched-filtering 
techniques~\citep{2026arXiv260527224T}, where the data is correlated with 
banks of \ac{CBC} waveform model parameter sets (hereafter referred to as 
\textit{templates}) that 
span a common parameter space across all pipelines, yielding the \ac{SNR} for 
each candidate. 
The template banks are designed to search for \acp{CBC} with redshifted 
primary 
masses $m_1^{\rm det} = (1+z) m_1$ between \MOneMIN and \MOneMAX, and redshifted 
secondary 
masses $m_2^{\rm det} = (1+z) m_2$ between \MTwoMIN and \MTwoMAX.
Here, $m_{1,2}^{\rm det}$ is the detector frame mass and $m_{1,2}$ is 
the source frame mass. 
The mass ratio, $\massratio = 
\masstwo/\massone$ ($\masstwo \leq \massone$) is restricted within 
\MassRatioMin--\MassRatioMax to maintain a computationally manageable number of 
templates in the template bank.
In the low mass regime considered for \ac{SSM} search, the inspiral signal is 
principally determined by a single combination of component masses, the (detector
frame) chirp mass:
\begin{equation}\label{eq:mc}
    \Mc^{\rm det} = (1+z)\Mc = \frac{(\massone \masstwo)^{3/5}}{(\massone + \masstwo)^{1/5}}.
\end{equation}
Mismatches between signal waveforms with different binary parameters vary much more
rapidly with differences in $\Mc^{\rm det}$ than with mass ratio or
component spins at constant $\Mc^{\rm det}$.  The sensitivity of search templates to
binary signals over a range of component masses and spins may thus be approximately
understood in terms of the range of $\Mc^{\rm det}$ covered. 

We include spin--spin and spin--orbit interaction effects assuming that the 
spin vectors of the two bodies are either parallel or anti-parallel to the orbital 
axis. 
The dimensionless spin magnitudes satisfy $\spincomponent \leq \MaxSpinHighMass$ for 
$m_{1,2}^{\rm det} \geq \SSMLowHighMassGap\,\Msun$ and 
$\spincomponent \leq \MaxSpinLowMass$ for $m_{1,2}^{\rm det} \leq \SSMLowHighMassGap\,\Msun$.
This search covers the same parameter space as the \ac{LVK} \ac{SSM} search 
performed on \ac{O3}~\citep{2022PhRvL.129f1104A, 2023MNRAS.524.5984L}
as well as \ac{O4a}~\citep{LIGOScientific:2026wxz}
data, making a trade-off between possible astrophysical targets and the 
computational cost of matched-filter searches in this low-mass region.

\GSTLAL, \MBTA, and \PYCBC search pipelines each produced template banks 
with comparable parameter space coverages. 
The three pipelines' template banks were constructed with a target minimal
match of $\GSTLALSSMMinMatch$, $\MinMatch$ and $\PyCBCSSMMinMatch$, for 
\GSTLAL, \MBTA, and \PYCBC, respectively. 
The chosen minimal match values are intended to balance the loss of 
astrophysical signals due to the discreteness of the bank  
(e.g.\ a maximum loss of $10\%$ for a minimal match of \MinMatch~\citep{1996PhRvD..53.6749O}) 
against computational cost considerations. 
The sensitivity bandwidth of \ac{LIGO} and Virgo extends down to approximately
\MinSensitiveFrequency\,Hz~\citep{2026arXiv260527223T}.
However, higher low-frequency cutoffs have been adopted in order to limit the 
computational cost of \ac{SSM} searches: a cutoff of \SSMFLOW is used by all 
pipelines when determining template bank placement, thus reducing the number of
templates to be processed; in addition, search pipelines impose low-frequency
cutoffs higher than \MinSensitiveFrequency\,Hz when performing matched filtering.  
This choice reduces computational cost per template, given that \ac{GW} signals from
low-mass \acp{CBC} can last for several minutes in the sensitive bandwidth of the
detectors.
Details of the \ac{SSM} template bank generation, search optimizations, and 
individual pipeline configurations are described in the following subsections.

\input{source__2.1-GstLAL.tex}

\input{source__2.2-MBTA.tex}

\input{source__2.3-PyCBC.tex}

\input{source__2.4-results.tex}

%% file: source__2.1-GstLAL.tex
\subsection{\GSTLAL Pipeline}\label{sec:2.1-gstlal}

\GSTLAL~\citep{2017PhRvD..95d2001M, 2019arXiv190108580S, 2021SoftX..1400680C, 2020PhRvD.101b2003H, 
2024PhRvD.109d2008E, 2023PhRvD.108d3004T, 2024PhRvD.109d4066S, 2023arXiv230607190R, 2023PhRvD.108h4032J, 
2025arXiv250606497J, 2025arXiv250523959J, Adhicary:2026vqu} 
is a GStreamer-based pipeline that searches for \ac{GW} signals from \acp{CBC} 
using matched-filtering techniques.
A concise overview of the methods developed and adopted for \ac{O4} analyses, 
including the latest archival search workflow employed in this work, is 
provided in~\citet{2026arXiv260527224T}.
Unlike in \ac{O3}~\citep{2022PhRvL.129f1104A, 2023MNRAS.524.5984L}, 
\GSTLAL uses a bin-dependent gating procedure to mitigate glitches instead of 
the iDQ algorithm~\citep{2024arXiv241204638H}.
The overview of the methods of \gwtcfour and \gwtcfive are provided 
in~\cite{2026ApJ..1004L..21A, 2026arXiv260527224T}, and~\citet{LIGOScientific:2026wxz} 
provides details on specific methods for \ac{SSM} searches.
The methodology, including the template bank~\citep{Hanna:2024tom}, used in the 
analyses for this paper remains nearly identical to that described in 
\citet{LIGOScientific:2026wxz} with differences described below. 

For \ac{O4} analyses, 
\GSTLAL generated the \ac{SSM} template bank using a geometric placement method, 
\MANIFOLD~\citep{2023PhRvD.108d2003H}.
The template bank used for the \ac{SSM} search of \ac{O4a} 
data~\citep{Hanna:2024tom} is inherited, as the search parameter space of 
interest remains identical and to avoid additional computational costs to 
remake a template bank.  
As in~\citet{LIGOScientific:2026wxz}, 
the \GSTLAL template bank consists of over \GstlalBankSize templates, 
the waveform approximant is 
\GstLALTemplateBankWaveformApproximant~\citep{2016PhRvD..93d4006H, 2016PhRvD..93d4007K}, the lower 
frequency cutoff is \GstlalFLOW or higher, the high frequency cutoff is 
\GstlalFHIGH, and a maximum template duration of $\sim\,$\GstlalMAXDUR is 
imposed~\citep{Hanna:2024tom}.
The limitation on template duration results in higher values of the 
low-frequency cutoff, given that the waveforms' \ac{GW} frequency increases 
monotonically with time. 

In \ac{O4} \GSTLAL searches, including the \ac{SSM} analysis, a number of 
enhancements are implemented in the algorithm compared to the \ac{O3} \ac{SSM} 
analysis~\citep{2023PhRvX..13d1039A}.
These include an improved ranking statistic~\citep{2023PhRvD.108d3004T}, a new 
singular value decomposition-based grouping method for parallelization~\citep{2024PhRvD.109d4066S}, and a 
template weighting derived from the \MANIFOLD population model~\citep{2023arXiv230607190R, 
2026ApJ..1004L..21A}. 
As in the \ac{O3} \ac{SSM} search, we continue to assume a uniform population 
model in the ranking statistic due to the absence of observational constraints 
on the underlying source population.
Data is match-filtered using the same frequency and waveform duration 
cutoffs applied in the template bank generation, with waveform models chosen 
based on the redshifted chirp mass $(1+z)\Mc$ where the 
\GSTLALLowMassWaveformApproximant 
approximant~\citep{2009PhRvD..80h4043B, Pan:2007nw, Boyle:2009dg} is used for 
$(1+z)\Mc \leq 1.73\,\Msun$, and the \GSTLALHighMassWaveformApproximant 
approximant~\citep{2009PhRvD..80h4043B, 2017PhRvD..95d4028B, 2020PhRvD.101l4040C} for 
$(1+z)\Mc > 1.73\,\Msun$.
We adopt different waveform models from that used for template placement in order to reduce computational costs.
After matched filtering, \ac{GW} candidates are 
ranked using a multi-component likelihood ratio ranking statistic~\citep{2023PhRvD.108d3004T, 2015arXiv150404632C}.
The component for multi-detector signal consistency~\citep{2020PhRvD.101b2003H} 
is evaluated over a frequency range of \GstlalFLOW to \GstlalDtDphiFHIGH. 
The ranking statistic is converted to \ac{FAR}, which quantifies the
significance of a candidate signal by comparing it to the distribution of 
noise triggers~\citep{2017PhRvD..95d2001M}.

The differences in methodology between what was used for \ac{O4a} \ac{SSM} 
analyses and the \ac{O4b} \ac{SSM} analyses are the following: the multi-detector 
signal consistency probability distribution, data calibration, and the detectors used 
for analyses. 
The multi-detector signal consistency probability distribution is a data product 
that assesses the signal probability of the \ac{GW} candidates based 
on the observed time difference and phase difference between detectors by downweighting
candidates with unphysical time or phase shifts. 
For the \ac{O4a} \ac{SSM} search, \GSTLAL used a high frequency cutoff of 
\GstlalDtDphiFHIGHOfoura for this data product, which was found to be inconsistent 
with the bandwidth used to whiten the templates.
While the discrepancy was shown to have negligible effects, the high frequency 
cutoff was corrected to \GstlalDtDphiFHIGH for \ac{O4b} \ac{SSM} analyses. 
Search pipelines process calibrated detector data. 
As described in~\citet{LIGOScientific:2026jgl}, there are periods of 
\ac{O4b} data that required additional calibration correction procedures that 
were not required for data in \ac{O4a}.
Hence, the \GSTLAL analysis modified its configurations to use the corrected 
data, which was publicly released~\citep{LIGOScientific:2026jgl}.  
Finally, \GSTLAL analyzed data from the two \ac{LIGO} detectors and the Virgo 
detector for \ac{O4b} analyses, as opposed to only the two \ac{LIGO} 
detectors in \ac{O4a}. 
The motivation was to increase the search's sensitivity by including Virgo 
data, which became available in \ac{O4b}.

%% file: source__2.2-MBTA.tex
\subsection{\MBTA Pipeline}\label{sec:2.2-mbta}

    The Multi-Band Template Analysis (\MBTA;~\citep{2012ApJ...760...12A, 2016CQGra..33q5012A, 2021CQGra..38i5004A, 2025CQGra..42j5009A}
    pipeline performs a matched-filter search for \acp{GW} from \acp{CBC} in data from the \ac{LIGO} and Virgo detectors, using a pre-computed template bank.
    To reduce computational cost, the \MBTA pipeline splits the matched-filtering process in the frequency domain into two bands, performing the filtering in each band before recombining the results~\citep{2026ApJ..1004L..21A}.

    Notable changes have been made since \ac{O4a}.
    The Virgo data has been added to the analysis.
    The template bank has been generated using the same parameters as for \ac{O4a}, but with a more recent \ac{PSD}.
    It comprises over \MBTABankSize templates covering the parameter space described above, and was generated using a geometric placement algorithm~\citep{2013PhRvD..87h2004B}.
    The low-frequency and high-frequency template banks have also been regenerated to allow filtering starting at \MBTAFilterFLOW.
    While the placement of the template bank is defined over the range \MBTABankFLOW--\MBTABankFHIGH, the filtering is performed over \MBTAFilterFLOW to \MBTAFilterFHIGH. The reduction of the filtering low-frequency cutoff yields an improvement in sensitivity, while halving the filtering high-frequency limit preserves such an improvement while ensuring a minimal impact on total computational cost.
    The waveforms used in constructing the template bank are computed using the \MBTAWaveformApproximant approximant~\citep{2009PhRvD..80h4043B, Pan:2007nw, Boyle:2009dg}; however for matched filtering,
    different waveform models are used depending on the total mass of the template.  \MBTASEARCHAPPROXLOWMASS~\citep{Isoyama:2020lls} is used when the total mass $\Mtot$ (= \massone + \masstwo) < 4.0 \Msun, whereas \MBTASEARCHAPPROXHIGHMASS~\citep{2021PhRvD.103l4060E} is used otherwise.

%% file: source__2.3-PyCBC.tex
\subsection{\PYCBC Pipeline}\label{sec:2.3-pycbc}

The \PYCBC pipeline~\citep{2012PhRvD..85l2006A, 2014PhRvD..90h2004D, 2016CQGra..33u5004U, 2017ApJ...849..118N, 2020PhRvD.102b2004D, pycbcsoft} 
performs a matched filtering analysis to detect \ac{GW} signals from \acp{CBC}.
The \PYCBC \ac{SSM} search employs the latest version of the pipeline, described in detail in~\citet{2026ApJ..1004L..21A}.
We made specific choices to tailor the analysis to the \ac{O4b} search for \ac{SSM} compact objects and to improve its sensitivity, as summarized below.

The ranking statistic is the same as in the \ac{O4a} \ac{SSM} search~\citep{LIGOScientific:2026wxz}.
The major improvement in O4b is that we analyze Virgo data in addition to \ac{LHO} and \ac{LLO}. 
As in \ac{O4a}, we incorporate single-detector triggers in the search following~\citep{2022CQGra..39u5012C}; 
the threshold for fitting the Virgo single trigger background is tuned with a representative chunk of data during \ac{O4b}.

For the template bank generation, the \PyCBCWaveformApproximant~\citep{2009PhRvD..80h4043B, Pan:2007nw, Boyle:2009dg} post-Newtonian approximant 
and a lower frequency cutoff of \PyCBCBankFLOW were adopted to compute the matches between templates.
We use the same geometric placement algorithm~\citep{2013PhRvD..87h2004B} as \ac{O4a}, though now considering a representative \ac{PSD} for \ac{LHO}  and \ac{LLO} during \ac{O4b}. 
This results in a bank with \PyCBCBankSize templates.

The pipeline starts filtering the data at \PyCBCFilterFLOW with an imposed maximum allowed template duration of \PyCBCMaxTempltDur which can lead to truncated templates.
The \PyCBCFilteringWaveformApproximant approximant~\citep{2009PhRvD..80h4043B, 2017PhRvD..95d4028B, 2020PhRvD.101l4040C} is used for matched filtering 
in templates with $\Mtot$ $> \PyCBCMtotThrforApprox$, while the \PyCBCWaveformApproximant approximant is used for $\Mtot \leq \PyCBCMtotThrforApprox$.

%% file: source__2.4-results.tex
\subsection{Search Results}
\label{sec:results}

\subsubsection{Summary}

\begin{figure*}
	\centering
	\includegraphics[width=1.0\textwidth]{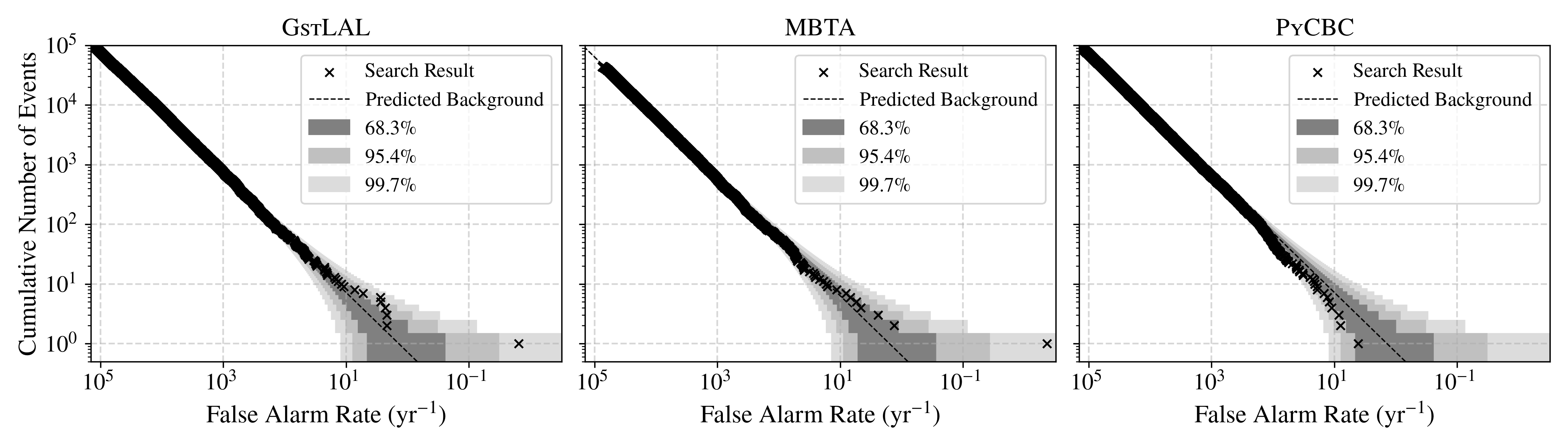}
	\caption{Cumulative number of \ac{SSM} candidates as a function of the \ac{FAR} for the \GSTLAL (left panel), \MBTA (middle panel), and \PYCBC (right panel) analyses.
		The black dashed line shows the predicted background, with the shaded regions indicating the 68\%, 95\%, and 99\% confidence intervals for the background count at any \ac{FAR} threshold, derived from a Poisson distribution given the total analyzed time of \ac{O4b} and the preceding engineering run.	Black crosses indicate the observed cumulative count from each pipeline.
		The low \ac{FAR} candidates seen in the \GSTLAL and \MBTA distributions correspond to \GSTLALFirstTriggerSEventName and \MBTAFirstTriggerSEventName, respectively.} 
	\label{fig:search_results_FAR_plots}
\end{figure*}
We report no significant \ac{GW} candidates from \ac{SSM} searches using the
\GSTLAL, \MBTA, and \PYCBC pipelines conducted on \ac{O4b} data. 
Figure~\ref{fig:search_results_FAR_plots}  
shows the number of candidates with estimated \ac{FAR}  
exceeding a given threshold, as a function of the threshold value.
The observed candidates are fully consistent with the Poisson statistical expectation, 
except for two outliers identified by the \GSTLAL pipeline on \GSTLALFirstTriggerTime 
and by the \MBTA pipeline on \MBTAFirstTriggerTime, discussed in detail below. 

Table~\ref{tab:triggers} lists candidates identified by the search pipelines with a \ac{FAR} 
below \SignificantOPA. 
The third most significant candidate, despite having a higher \ac{SNR} than the two 
outliers, is ranked lower due to the reduced sensitivity of the contributing detectors
(\ac{LHO} and {Virgo} rather than \ac{LHO} and \ac{LLO}).  As its \ac{FAR} is consistent
with noise origin given the observing time of \ac{O4b}, we do not investigate this candidate
further. 
\begin{table*}[tbp]
  \setlength{\tabcolsep}{3pt}
  \renewcommand{\arraystretch}{1.2}
  \caption{\label{tab:triggers}Candidates recovered with a \ac{FAR} below \SignificantOPA.}
  \begin{ruledtabular}
    \begin{tabular}{ c  c  c  c  c  c  c  c  c  c }
    \ac{FAR} & Pipeline & Coalescence time & $\massone^{\rm det}$ & $\masstwo^{\rm det}$ & $\chi_{1z}$ & $\chi_{2z}$ & Inst.\ operating & Inst.\ contributing  & Network \ac{SNR} \\
    ($\mathrm{yr^{-1}}$) &&& (\Msun) & (\Msun) &&&& \\
    \hline
    \expandafter\MBTATableRow\MBTATrigOneData
    \expandafter\GSTLALTableRow\GSTLALTrigOneData
    \expandafter\MBTATableRow\MBTATrigTwoData

    \end{tabular}
  \end{ruledtabular}
\tablecomments{For each candidate we report the redshifted primary (1) and secondary (2) masses $\masscomponent^{\rm det}$ and the dimensionless spin components projected along the orbital axis $\chi_{1,2z}$, for the highest ranked template.
The detectors operating at the time of the candidate are listed under ``Inst. operating'', while detectors where the \ac{SNR} is high enough to contribute to the estimated significance are listed under ``Inst. contributing'', where H (L, V) denotes \acl{LHO} (\acl{LLO}, Virgo).
The minimum \ac{SNR} thresholds for a detector to contribute to candidate significance for \GSTLAL and \MBTA are \GSTLALSNRThreshold and \MBTASNRThreshold, respectively. 
The network \ac{SNR} is the quadrature sum of \acp{SNR} for all detectors operating at the time of the candidate for \MBTA, or only for detectors with \ac{SNR} above \GSTLALSNRThreshold for \GSTLAL. 
}
\end{table*}

\paragraph{Candidate significance and trials factors}
Considering the duration of \ac{O4b} data, the \acp{FAR} reported for the most
significant \GSTLAL and \MBTA candidates correspond to a \ac{FAP} (or p-value) of 
\GSTLALFirstTriggerGSTLALSSMOfflineTriggerFAP and \MBTAFirstTriggerFAP, respectively. 
Thus, neither of these candidates surpasses a $3\,\sigma$ threshold, and we do not
consider them as providing evidence for detection.  We present follow-up studies on
the two outlier candidates below. 

The significance reported for individual offline \ac{SSM} search pipelines over
\ac{O4b} does not incorporate trials factors, which account for the fact that 
multiple distinct search results have been reported using \ac{aLIGO} and \ac{aVirgo}
data, and every search has a finite probability of producing a false positive. 
To attempt to quantify this effect, we consider the \ac{LVK}'s \ac{SSM} searches
conducted from \ac{O1} through \ac{O4b}.  
The numbers of pipelines participating 
for each observing run are as follows: 
\begin{itemize}
	\item \ac{O1}: \NumberOfPipelinesOOne (\NameOfPipelinesOOne)
	\item \ac{O2}: \NumberOfPipelinesOTwo (\NameOfPipelinesOTwo)
	\item \ac{O3a}: \NumberOfPipelinesOThreeA (\NameOfPipelinesOThreeA)
	\item \ac{O3b}: \NumberOfPipelinesOThreeB (\NameOfPipelinesOThreeB)
	\item \ac{O4a}: \NumberOfPipelinesOFourA (\NameOfPipelinesOFourA)
	\item \ac{O4b}: \NumberOfPipelinesOFourB (\NameOfPipelinesOFourB)
\end{itemize}
resulting in \TrialsFactorAcrossObservingRuns separate search results.
While the effect of these multiple searches on the significance threshold
required to claim evidence for observation of an astrophysical signal is difficult
to quantify exactly, if we consider events due to detector noise in different 
pipelines to be independent to a good approximation, the \ac{FAP} of a candidate 
will increase by around an order of magnitude when considering trials factors. 

\subsubsection{\GSTLAL \ac{SSM} Outlier Candidate}

The outlier candidate identified by the \GSTLAL pipeline corresponds to 
\GSTLALFirstTriggerSEventName~\citep{GCN36236} in the \ac{LVK} \GRACEDB~\citep{GraceDB2014}.
All three detectors were operating at the time of the candidate but triggers were 
produced above the pipeline's \ac{SNR} threshold only in \ac{LLO}
and \ac{LHO}. 

This candidate was initially identified by the low-latency stellar-mass (broad
\ac{CBC} parameter space) search in the \GSTLAL and \PYCBC pipelines, with 
\acp{FAR} of 
\GSTLALFirstTriggerGSTLALAllSkyOnlineTriggerFAR and \GSTLALFirstTriggerPYCBCAllSkyOnlineTriggerFAR, respectively. 
Both pipelines identified the candidate in templates with masses consistent with 
a \ac{NSBH} source binary, with a (redshifted) chirp mass of 
\GSTLALFirstTriggerOnlineMchirp and a effective inspiral spin $\chieff \equiv 
(\massone\chi_{1z} + \masstwo\chi_{2z})/(\massone+\masstwo)$ of \GSTLALFirstTriggerOnlineChieff. 
The \MBTA pipeline's low-latency stellar-mass search did not detect this candidate.

At the time of the candidate, the \GSTLAL and \MBTA pipelines were operating  
low-latency \ac{SSM} searches~\citep{Hanna:2024tom, 2025CQGra..42j5009A}.
However, these searches did not identify candidates for \GSTLALFirstTriggerSEventName. 
The \GSTLAL low-latency \ac{SSM} template bank's spins are limited to 
$|\spincomponent| \leq \MaxGSTLALOnlineSSMSpin$~\citep{Hanna:2024tom}, 
hence, this bank did not cover the preferred template parameters for \GSTLALFirstTriggerSEventName. 

Further analyses of this candidate in the \GSTLAL stellar-mass search including 
information from subsequent data in the background model reduced its significance; 
the \ac{FAR} in the \GSTLAL stellar-mass search is now estimated to be 
\GSTLALFirstTriggerGSTLALAllSkyOnlineTriggerFARUpdated, 
comparable with the \PYCBC pipeline's stellar-mass search~\citep{GCN36812}.
While \GSTLALFirstTriggerSEventName is included in \thisgwtc~\citep{2026arXiv260527225T}, 
it is no longer a significant candidate.  

However, the \GSTLAL \ac{SSM} offline search of \ac{O4b} data reported here identifies a
candidate corresponding to \GSTLALFirstTriggerSEventName with a \ac{FAR} of 
\GSTLALFirstTriggerGSTLALSSMOfflineTriggerFAR. 
Neither the \MBTA nor the \PYCBC offline \ac{SSM} searches identify a corresponding 
candidate for \GSTLALFirstTriggerSEventName at a \ac{FAR} inconsistent with noise.

Data quality checks performed for \GSTLALFirstTriggerSEventName in low latency
indicated that data from both \ac{LLO} and \ac{LHO} were affected by
non-stationary noise, 
but not at a level sufficient to justify retracting the alert. Additionally,
the trigger identified in \ac{LHO} data is well within the distribution of \SNR
and chi-squared values for noise triggers. 
Follow-up Bayesian inference analyses, under the hypothesis of an astrophysical signal in Gaussian noise, infer a chirp mass consistent with the search pipeline value from coherent analysis of data from all three detectors. 
However, both \Ac{LHO} data and Virgo data, when analyzed on their own, are consistent with a vanishing signal amplitude for this range of chirp mass. 
While such behaviour is not impossible for an astrophysical signal, overall these followup analyses do not change our assessment that the \GSTLAL candidate is consistent with a noise origin.

\subsubsection{\MBTA \ac{SSM} Outlier Candidate}

The outlier candidate identified by the \MBTA pipeline 
corresponds to \MBTAFirstTriggerSEventName.
As with \GSTLALFirstTriggerSEventName, all three detectors were operating 
at the time of the candidate but triggers were produced above the pipeline's \ac{SNR}
threshold only in \ac{LLO} and \ac{LHO}. 

This candidate was initially identified as \MBTAFirstTriggerSEventName by the 
low-latency \ac{SSM} search in the \MBTA and \GSTLAL pipelines, with a \ac{FAR} of
\MBTAFirstTriggerMBTASSMOnlineTriggerFAR and \MBTAFirstTriggerGSTLALSSMOnlineTriggerFAR, 
respectively.
The \PYCBC pipeline's low-latency stellar-mass search did not detect
this candidate; note that the \PYCBC pipeline does not operate a low-latency \ac{SSM}
search. 
Since \MBTAFirstTriggerSEventName was identified as a low-significance candidate, 
a public alert was not issued, and neither manual checks for data quality nor
low-latency \ac{PE} were performed. 

As shown in Table~\ref{tab:triggers}, the \MBTA \ac{SSM} offline search of \ac{O4b} data 
identifies a candidate corresponding to \MBTAFirstTriggerSEventName with a 
\ac{FAR} of \MBTAFirstTriggerFAR, equivalent to a significance of \MBTAFirstTriggerSignificance. 
(As mentioned previously, this value does not account for trials factors.) 
The \GSTLAL and \PYCBC offline \ac{SSM} searches both identify a candidate corresponding to
\MBTAFirstTriggerSEventName, but with \acp{FAR} consistent with noise: \MBTAFirstTriggerGSTLALSSMOfflineTriggerFAR and \MBTAFirstTriggerPyCBCSSMOfflineTriggerFAR for \GSTLAL and \PYCBC, respectively. The probability for
the \ac{FAR} of an astrophysical signal to be higher than \MBTAFirstTriggerFARDiscrepancyCutHigh for \GSTLAL and \PYCBC, but lower than \MBTAFirstTriggerFARDiscrepancyCutLow for \MBTA, was estimated via analysis of simulated signals (\emph{injections}, see Section~\ref{sec:3-search_sensitivity} below) to be of the order of \MBTAFirstTriggerFARdiscrepancy.

Further investigations confirmed that the \MBTA noise model used to estimate the \ac{FAR} for this candidate was stable across the observing period, and that the candidate was not affected by data quality issues.
The \ac{SNR} evolution of the trigger when lowering the matched filter low frequency cutoff was also investigated, finding a decrease of the \ac{SNR}, and consequently of the significance, between \MBTAFilterFLOW and \MBTAFirstTriggerTestFlow cutoffs.
A decrease of the \ac{SNR} is also observed when performing the matched filtering over a single frequency band that covers the range of the usual two frequency bands in MBTA.
Further checks on injections indicate that this behavior, while possible for a low-amplitude astrophysical signal, was approximately twice as likely to be caused by a noise fluctuation.

%% file: source__3-search_sensitivity.tex
\section{Search Sensitivity}\label{sec:3-search_sensitivity}

In this section we quantify the sensitivity of our searches over \ac{O4a} and \ac{O4b} data to \ac{GW} signals from \ac{SSM} \acp{CBC}.
This enables us to set upper limits on \ac{SSM} merger rates, as described in Section~\ref{sec:4-rates}. 
Due to the changing sensitivity across the observing run and the complex nature of the detection process, we determine this sensitivity empirically via the injection procedure laid out in~\citet{2026ApJ..1004L..21A, 2025PhRvD.112j2001E}.
We simulate the \ac{GW} strain for large numbers of signals and add them to detector strain data, and pipelines perform the search described in Section~\ref{sec:2-search} on this augmented data. 
Our sensitivity to a possible population of \ac{SSM} sources depends on their distribution over masses and spins. 
Therefore, we provide estimates of sensitivity to \ac{SSM} sources as a function of an assumed population. 
Re-weighting a single set of results from our injection recovery analysis via importance sampling~\citep{2018CQGra..35n5009T} enables us to determine the sensitivity to multiple population models.

\subsection{Estimation methods}\label{sec:3.1-estimation_method}

Search sensitivity for a given assumed population distribution $\mathbf{\Lambda}$, with a merger rate which we assume constant over comoving volume and time, is described by the sensitive hypervolume, $\langle VT\rangle$.  This is an effective spacetime volume over which the search would detect mergers with a significance above some chosen detection threshold.  It may be expressed as 
\begin{equation}\label{eq:VT_from_xi}
    \langle VT \rangle(\mathbf{\Lambda}) = \xi(\mathbf{\Lambda}) VT_{\rm total}\,,
\end{equation}
where $\xi(\mathbf{\Lambda})$ is the fraction of detectable mergers, and $VT_{\rm total}$ denotes the total volume--time accessible to the analysis, given by the comoving volume--time enclosed out to the most distant source included in the population, over the duration $T_{\rm obs}$ of the data searched. 
For injections generated up to a maximum redshift $z_{\rm max}$, we have 
\begin{align}\label{eq:VT_total}
    VT_{\rm total} = T_{\rm obs} \int^{z_{\rm max}}_0 \frac{1}{1+z} \frac{\mathrm{d}V_{\mathrm{c}}}{\mathrm{d}z} \mathrm{d}z\,, 
\end{align}
where the $1/(1+z)$ factor accounts for time dilation between the source frame and the detector frame, and $\mathrm{d}V_{\rm c}/\mathrm{d}z$ is the differential element of comoving volume over redshift.
The detectable fraction $\xi(\mathbf{\Lambda})$ is defined as
\begin{equation}\label{eq:pdet}
    \xi(\mathbf{\Lambda}) = \int p_{\rm det}(d) p(d|\boldsymbol{\theta}) 
    p(\boldsymbol{\theta}|\mathbf{\Lambda}) \,\mathrm{d}d \,\mathrm{d}\boldsymbol{\theta}\,,
\end{equation}
where $p(d|\boldsymbol{\theta})$ denotes the probability of obtaining data $d$ in the presence of a \ac{GW} signal characterized by source parameters $\boldsymbol{\theta}$, and $p(\boldsymbol{\theta}|\boldsymbol{\Lambda})$ denotes the probability distribution of these parameters conditioned on the population hyperparameters $\boldsymbol{\Lambda}$. 
Finally, $p_{\rm det}(d)$ is the probability that our search identifies a candidate above the given detection threshold, when given data $d$ as input; as our search pipelines process data deterministically, this takes values of either $0$ or $1$.  

We estimate $\xi(\boldsymbol{\Lambda})$ for a given population model $\boldsymbol{\Lambda}$ via importance sampling~\citep{2018CQGra..35n5009T,Essick:2022ojx}.  We perform a Monte Carlo sum over the subset of recovered injections $\{\theta\}_{\mathrm{found}}$, applying appropriate importance sampling weights for the desired population distribution: 
\begin{equation}
    \xi(\mathbf{\Lambda}) \simeq \frac{1}{N_{\rm inj}} \sum_{\boldsymbol\theta\in\{{\boldsymbol\theta}\}_{\rm found}} \frac{p(\boldsymbol{\theta}|\mathbf{\Lambda})}{p(\boldsymbol{\theta}|\mathbf{\Lambda}_0)},
    \label{eq:MC_efficiency}
\end{equation}
where $N_{\rm inj}$ is the total number of injections initially generated, $p(\boldsymbol{\theta}| \boldsymbol{\Lambda}_0)$ is the distribution of source parameters under the fiducial population model $\boldsymbol{\Lambda}_0$ used to generate injections, and  
$p(\boldsymbol{\theta}| \boldsymbol{\Lambda})$ is the corresponding parameter distribution for the desired population model $\boldsymbol{\Lambda}$ for which we want to estimate sensitivity.  The uncertainty in the estimated $\langle VT \rangle$ propagates directly from the variance in this Monte Carlo sum~\citep{2019RNAAS...3...66F} via Equation~\eqref{eq:VT_from_xi}.

\subsection{Injections}\label{sec:3.2-injections}

To determine the sensitivity to populations of \acp{CBC} which include at least one \ac{SSM} \ac{BH} or \ac{NS} component in \ac{O4b}, we define two fiducial populations from which we draw simulated source parameters when constructing the injection sets used by the searches.
The injections were produced using \acp{PSD} derived from \ac{LHO}, \ac{LLO} and Virgo data throughout \ac{O4b}, and were added to the \ac{O4b} strain data for searches.
These populations are summarized in Table~\ref{tab:injection_population}, and are similar to those considered in the \ac{O4a} analysis of~\citet{LIGOScientific:2026wxz}, differing only in the aspects highlighted below. 
\newcommand{\multrowc}[1]{\begin{tabular}{@{}c} #1 \end{tabular}} 
\begin{table*}[tbp]
    \setlength{\tabcolsep}{20pt}
    \renewcommand{\arraystretch}{1.1}
    \caption{\label{tab:injection_population} Fiducial populations of simulated \ac{BBH} and \ac{BNS} sources used for estimating search sensitivities.}
    \begin{ruledtabular}
    \begin{tabular}{ c  c  c }
        & BBH & BNS \\
        \hline 
        \textsc{Mass Model}     & \multrowc{ $\masscomponent^{\alpha}$, $\alpha=\InjMassPowerLawAlpha$ \\ $\massone \in [\MinPrimaryMassBH, \MaxPrimaryMassBH]$ \\ $\masstwo \in [\MinSecondaryMassBH, \MaxSecondaryMassBH]$ } &  
                                  \multrowc{ $\masscomponent^{\alpha}$, $\alpha=\InjMassPowerLawAlpha$ \\ $\massone \in [\MinPrimaryMassNS, \MaxPrimaryMassNS]$ \\ $\masstwo \in [\MinSecondaryMassNS, \MaxSecondaryMassNS]$ } \\
        \hline
                                            & \multicolumn{2}{c}{$w\mathcal{N}(\mu,\sigma) + (1-w)\mathrm{Beta}(\alpha,\beta)$} \\
        \textsc{Spin Magnitude Model}       & \multicolumn{2}{c}{$\mu=\NormalDistMu, \sigma=\NormalDistSigma, \alpha = \BetaDistAlpha, \beta = \BetaDistBeta, w=\DistributionWeight$} \\
                                            & \multicolumn{2}{c}{$\spincomponent  \in [\MinSpinMag,\MaxSpinMag]$} \\
        \hline
        \textsc{Waveform Model} & \IMRPHENOMXPHM & \IMRPHENOMXPNRTIDALVTHREE~\citep{Abac:2023ujg} \\
        \hline
        \textsc{Equation of State} & Not used & \GPPVADDtwo \\
    \end{tabular}
    \end{ruledtabular}
\tablecomments{Source parameters are drawn from the given distributions for (source-frame) component masses $\masscomponent$, and dimensionless spin magnitudes \spincomponent, to generate simulated signals for searching augmented \ac{O4b} data.
We assume isotropic spin directions, i.e.\ uniform in the cosines of the angles between the orbital angular momentum vector and the component spin vectors.
The sources are distributed uniformly in comoving volume and time up to a maximum redshift $z_{\rm max} = \MaxRedshiftInjBH\,(\MaxRedshiftInjNS)$ for \ac{BBH} (\ac{BNS}).
These fiducial populations are denoted by $\mathbf{\Lambda_0}$ in Equation~\eqref{eq:MC_efficiency}.}
\end{table*}

The maximum primary mass for the \ac{BNS} population is set to $\MaxPrimaryMassNS$ (as compared to \MaxPrimaryMassNSOfoura adopted in \ac{O4a}), to narrowly include the maximum \ac{NS} mass of approximately \EOSMaxMassNS, which is allowed by the \ac{EOS} used to produce the injections in both \ac{O4a} and \ac{O4b}.
We adopt the $\GPPVADDtwo$ \ac{EOS}, which is closely related to DD2 and represents one of the stiffest \ac{EOS} still consistent with GW170817 multimessenger constraints~\citep{Radice:2017lry}, making $\MaxPrimaryMassNS$ a physically reasonable upper limit.
Other \ac{EOS} considered in~\cite{LIGOScientific:2018cki} are either too soft or excluded by multimessenger observations of GW170817; meanwhile, studies including NICER pulsar X-ray observations do not require a higher maximum mass~\citep{2021ApJ...918L..29R}.

The simulated population, for both \ac{BBH} and \ac{BNS}, has isotropically distributed spin orientations and is distributed uniformly in comoving volume out to a given maximum redshift $z_{\max}$.
For \ac{BNS} we set $z_{\max}=\MaxRedshiftInjNS$, which corresponds to a total hypervolume of $VT_{\rm total}=\TotalInjVTbns$ while for \ac{BBH} we use $z_{\max}=\MaxRedshiftInjBH$, corresponding to $VT_{\rm total}=\TotalInjVTbbh$. 
Compared to \ac{O4a} where a limit $z_{\max}=\MaxRedshiftInjNS$ was applied to all injections, here we extend the range for \ac{BBH}, motivated by the improvement in LIGO sensitivity and the higher network \ac{SNR} with Virgo joining the observing run~\citep{2026arXiv260527223T}.

We injected a total of $\TotalInjections$ simulated signals; \ac{BBH} signals are injected every $\InjSpacing$, and \ac{BNS} signals are injected on the same cadence, but offset by $\MBTAInjSpacing$ relative to the \ac{BBH} injections. These spacings are intended to avoid significant bias of the strain data \ac{PSD} caused by the presence of a high total signal rate.

\subsection{Combined \ac{O4a} and \ac{O4b} sensitivity estimates}

In this study we introduce an estimation of search sensitivity to \ac{SSM} binaries over the combined runs \ac{O4a} and \ac{O4b}.
This combined estimate uses an increased total observing time $T_{\rm obs}$, while consistently accounting for different search efficiencies $\xi(\mathbf{\Lambda})$ between runs.
We follow \cite{2021RNAAS...5..220E} to construct a mixture model for both runs, using the \ac{O4a} injections detailed in \cite{LIGOScientific:2026wxz} and \ac{O4b} injections described above.
The mixture unifies the two injection sets, assigning individual injection weights based on the rate of Monte Carlo sample generation in any given piece of data.  This weighting ensures uniformity over detector-frame time throughout the observing period, even in disjoint runs; the resulting weights are incorporated in the term $p(\boldsymbol{\theta}|\mathbf{\Lambda}_0)$ for Equation~\eqref{eq:MC_efficiency}~\citep{2021RNAAS...5..220E,2025PhRvD.112j2001E}.
The procedure described in Section~\ref{sec:3.1-estimation_method} is then applied to the mixture model set to obtain sensitivity estimates for the combined runs.
For runs that are disjoint in time, this combined estimate of $\langle VT \rangle$ for a given population is equal to a sum of estimated sensitivities over separate runs. However, the mixture model method has the advantage of being more generally applicable and valid even in cases of overlapping injection distributions (including overlapping in time).  Hereafter, combined sensitivity estimates for both runs will be labelled as \Ofourab.

The results and analyses presented in the remainder of this study adopt different detection thresholds for sensitivity estimation corresponding to distinct inference methods. 
Table~\ref{tab:FARThreshold_per_analysis} summarizes these detection thresholds, and the corresponding figures, for later reference. 
\newcommand{\multrowl}[1]{\begin{tabular}{@{}l} #1 \end{tabular}} 
\begin{table*}[tbp]
    \centering
    \setlength{\tabcolsep}{18pt}
    \renewcommand{\arraystretch}{1.1}
    \caption{\label{tab:FARThreshold_per_analysis} Significance thresholds used for search sensitivity estimates in downstream analyses, and the corresponding figures.}
    \begin{ruledtabular}
    \begin{tabular}{c c c}
        Analysis & Detection threshold & Figures \\
        \hline
        Low-spin \ac{SSM} \ac{BBH} $\langle VT \rangle$ estimates & \ac{FAR} $= 1\,\mathrm{yr^{-1}}$ & \ref{fig:o4ab_vt}, \ref{fig:o4a_o4b_o4ab_per_pipeline}, \ref{fig:o4ab_m1m2}, \ref{fig:o4ab_ratio_spins} \\
        High-spin \ac{SSM} \ac{BBH} $\langle VT \rangle$ estimates & \ac{FAR} $= 1\,\mathrm{yr^{-1}}$ & \ref{fig:o4ab_ratio_spins} \\
        \ac{SSM} \ac{BNS} $\langle VT \rangle$ estimates & \ac{FAR} $= 1\,\mathrm{yr^{-1}}$ & \ref{fig:o4ab_bns_vt} \\
        \hline
        \multrowc{Low-spin \ac{SSM} \ac{BBH}\\frequentist merger rate limits} & \multrowc{Per-pipeline loudest event\\over \ac{O4a} and \ac{O4b}} & \ref{fig:o4a_o4ab_rates}, \ref{fig:o4ab_rates_m1m2} \\
        \multrowc{\ac{SSM} \ac{BNS} frequentist\\merger rate limits} & \multrowc{Per-pipeline loudest event\\over \ac{O4a} and \ac{O4b}} & \ref{fig:o4a_o4ab_bns_rates} \\
        \multrowc{\ac{PBH} constraints derived from\\low-spin \ac{SSM} \ac{BBH} rate limits} & \multrowc{Per-pipeline loudest event\\ over \ac{O4a} and \ac{O4b}} & \ref{fig:fPBH_O4b_withLB}, \ref{fig:FPBH_m1m2_O4ab_EB} \\
        \hline
        Bayesian constraints on \ac{DBH} model & \ac{FAR} $= 0.01\,\mathrm{yr^{-1}}$  & \ref{fig:atomic_dark_matter_f_vs_mmin} \\  
    \end{tabular}
    \end{ruledtabular}
\end{table*}

\input{source__3.1-SSM_BBH.tex}

\input{source__3.2-SSM_BNS.tex}

%% file: source__3.1-SSM_BBH.tex
\subsection{Sub-solar mass BBH systems}\label{sec:bbh_vt}

We report sensitivity estimates, $\langle VT \rangle$, for \ac{BBH} systems with at least one \ac{SSM} component recovered by the \GSTLAL, \MBTA and \PYCBC pipelines.
We expect search sensitivity to be primarily a function of the binary chirp mass~\citep{2014LRR....17....2B}
as the \ac{GW} frequency evolution and emitted flux during binary inspiral are determined, to first approximation, only by the value of $\Mc$~\citep{1994PhRvD..49.2658C}. 
Considering both the masses covered by the search template banks in the detector frame, and the maximum redshift of $\MaxRedshiftInjBH$ reached by our injection campaign, we consider a chirp mass space covering the range $\MinMcVT \leq \Mc \leq \MaxMcVT$; we then divide this space into equally-spaced bins, treating each bin as a separate population.
Additionally, we reject binaries with $\massratio < \MinMassRatioVTpop$ 
for consistency with previous estimates in~\cite{2022PhRvL.129f1104A, 2023MNRAS.524.5984L, LIGOScientific:2026wxz}.
Note that since search pipelines can have sensitivities to injections that lie outside the parameter spaces covered by their template banks, we assess the sensitivities using a cutoff at $\massratio = \MinMassRatioVTpop$ while the template banks are limited to $\massratio$ within \MassRatioMin--\MassRatioMax. 

In this study we estimate sensitivity for low-spin binaries, as previously in~\citet{LIGOScientific:2026wxz}, and also for high-spin \ac{SSM} binaries. 
Low-spin SSM binaries correspond to the simplest scenario of \acp{PBH} formed by gravitational collapse of overdensities that enter the horizon in the radiation era, which are expected to have zero or negligible spin at formation~\citep{Garcia-Bellido:2017fdg, Chiba:2017rvs, 2020JCAP...03..017M, 2019JCAP...05..018D}. 
However, several mechanisms could increase the spin of one or both components of the coalescing binaries.
In close hyperbolic encounters, \acp{PBH} can acquire a relatively mild spin of order $\chi\sim 0.2$~\citep{Jaraba:2021ces}.
In sufficiently dense clusters of \acp{PBH}, the spin can reach $\HighSpinPBH$ due to this mechanism only if significantly unequal mass \acp{BH}
participate in multiple close hyperbolic encounters.
Alternatively, if the \acp{PBH} inhabit a sufficiently gas-rich environment for accretion disks to be formed, the transfer of angular momentum 
due to gas accretion can increase substantially the spin magnitude~\citep{2020JCAP...04..052D}, although this mechanism needs to be confirmed with simulations of \ac{SSM} \acp{PBH}.

For the low-spin case, we choose a uniform distribution in spin magnitudes limited to $\spincomponent \leq \MaxLowSpinMagVTpop$, consistent with previous reported results~\citep{2022PhRvL.129f1104A, 2023MNRAS.524.5984L, LIGOScientific:2026wxz}. 
Then, for the high-spin case we choose a spin magnitude distribution uniform in $\MaxLowSpinMagVTpop \leq \spincomponent \leq \MaxHighSpinMagVTpop$: the upper limit here is motivated by upper bounds in the literature~\citep{2020JCAP...04..052D, Jaraba:2021ces}, and in order to avoid an excess of variance in the estimate of $\langle VT \rangle$~\citep{2019RNAAS...3...66F} caused by low support of the injected $\spincomponent$ distribution beyond $\MaxHighSpinMagVTpop$.
Spin orientations are considered as isotropically distributed for both low- and high-spin cases. 

In addition to reporting sensitivity estimates in bins of chirp mass, we also report sensitivity as a function of component masses $\masscomponent$.
Similarly to the chirp mass case, we split the space of component masses into equally spaced bins covering the range $\MinMOneVT \leq \massone \leq \MaxMOneVT$ and $\MinMTwoVT \leq \masstwo \leq \MaxMTwoVT$, with the mass ratio constraint $q \geq \MinMassRatioVTpop$, and treat each grid bin as a separate population, described by a log-uniform distribution limited by the bin bounds. 
For component mass sensitivity estimates, we only consider the low-spin case with a uniform magnitude distribution $\spincomponent \leq \MaxLowSpinMagVTpop$ and isotropically distributed spin orientations.

\begin{figure}[h]
    \centering
        \includegraphics[width=1.0\columnwidth]{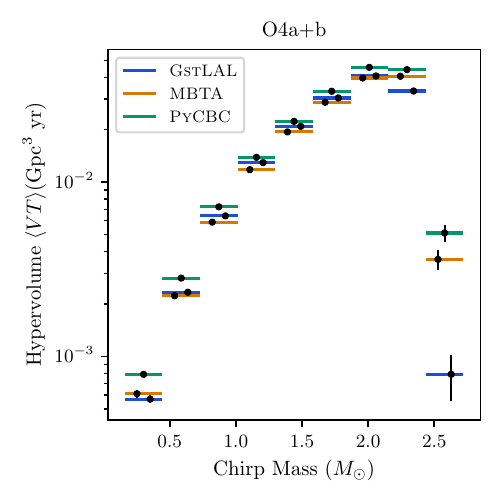}
    \caption{
        Low spin \ac{BBH} population sensitive hypervolume estimates at a \ac{FAR} $\BBHVTFARThreshold$ threshold for each pipeline over the combined \Ofourab runs, computed in chirp mass bins.
        Vertical bars show Monte Carlo sampling standard deviations.
        }
    \label{fig:o4ab_vt}
\end{figure}

We present sensitivity estimates for \ac{SSM} \ac{BBH} at a reference significance threshold of \ac{FAR} $\BBHVTFARThreshold$ (in contrast to~\citet{LIGOScientific:2026wxz}, where a loudest event threshold was used).
This uniform choice follows~\citet{2026arXiv260527225T} and facilitates one-to-one comparisons between pipelines and/or between observing periods.
Figure~\ref{fig:o4ab_vt} shows the combined estimates for low-spin \acp{BBH}, with similar performance between different search pipelines. 
Larger differences in sensitivity between \PYCBC and the other two pipelines seen at low chirp mass are due to a lower starting frequency for filtering (see Section~\ref{sec:2.3-pycbc}), which affects the power recovered from long duration signals characteristic of low-mass binaries.

\begin{figure*}
    \centering
    \includegraphics[width=1.0\textwidth]{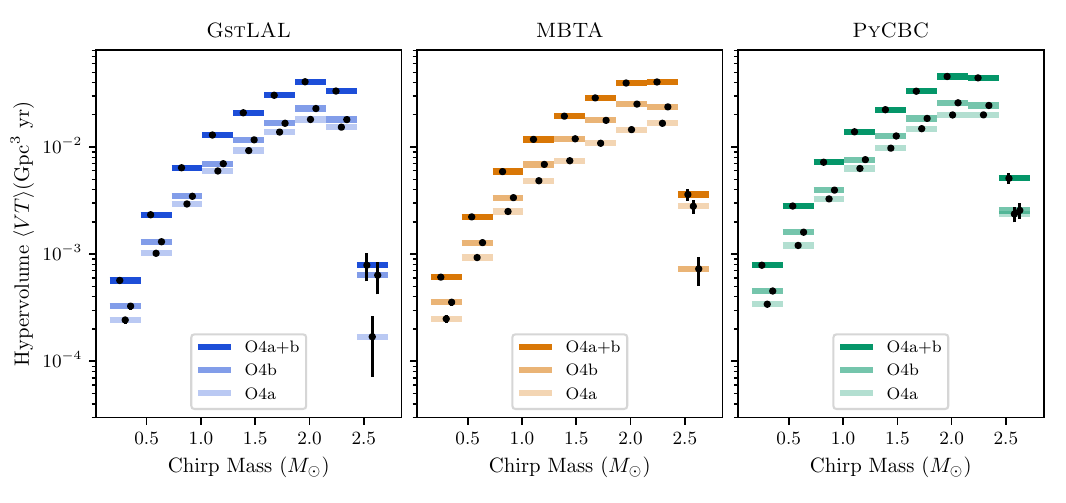}
    \caption{
        Low spin \ac{BBH} population sensitive hypervolume estimates at a \ac{FAR} $\BBHVTFARThreshold$ threshold for \ac{O4a}, \ac{O4b} and combined runs \Ofourab, displayed for each pipeline in chirp mass bins.
        Dimmer solid lines represent \ac{O4a} while subsequent runs are shown as darker lines.
        Vertical bars show Monte Carlo sampling standard deviations.
        }
    \label{fig:o4a_o4b_o4ab_per_pipeline}
\end{figure*}

We also show a comparison of sensitive $\langle VT \rangle$ in Figure~\ref{fig:o4a_o4b_o4ab_per_pipeline} for low-spin \acp{BBH} during \ac{O4a}, \ac{O4b} and the combination of both runs.
Increased sensitivity in \ac{O4b} is due to a combination of increased observing time from $\FourthAObsAnalysisTime$ (\ac{O4a}) to $\FourthBObsAnalysisTime$ (\ac{O4b}) translated in a $\approx\TobsIncrease$ gain; improved LIGO detector sensitivity, and the resumption of Virgo observations after \ac{O4a}; and improvements in search pipeline methods which enhance detection efficiency by up to $\EfficiencyIncrease$.
In particular, the \MBTA sensitivity increases for \ac{O4b} as compared to \ac{O4a} reaching a $\maxIncreaseAtoBMBTA$ improvement in higher chirp mass bins, which we can attribute to search improvements described in Section~\ref{sec:2.2-mbta}. On the other side, \GSTLAL and \PYCBC increase at most $\maxIncreaseAtoBGstLALPyCBC$. For the combined \Ofourab runs, $T_{\rm obs}$ increases by a factor $\increaseTobsABtoA$ with respect to \ac{O4a}.
Together with improvements to the detectors' sensitivities and the addition of the Virgo detector, the sensitivity gain is nearly a factor of $\increaseVtABtoA$.

Uncertainties in $\langle VT \rangle$ are propagated from the Monte Carlo sum standard deviation, as discussed in Section~\ref{sec:3-search_sensitivity}.
The great majority of injected signals in the last chirp mass bin, $\lastMcbinmin\text{--}\lastMcbinmax\,\Msun$, have a chirp mass above the range covered by the pipelines' template banks, leading to a much reduced sensitivity.
As the estimated sensitivity in this bin is subject to large statistical uncertainties, as follows from~\citet{2019RNAAS...3...66F}, we display the estimate here for completeness but we do not proceed to derive astrophysical limits for the last bin. 

The \ac{O4a} estimates~\citep{LIGOScientific:2026wxz} in Figure~\ref{fig:o4a_o4b_o4ab_per_pipeline} used loudest-event \ac{FAR} thresholds of $\FourthAGSTLALLoudestTriggerFAR$, $\FourthAMBTALoudestTriggerFAR$ and $\FourthAPYCBCLoudestTriggerFAR$ for \GSTLAL, \MBTA and \PYCBC respectively, while here we report $\langle VT \rangle$ at a $\BBHVTFARThreshold$ \ac{FAR} threshold.
This caveats a direct comparison, since varying the significance threshold will change the Monte Carlo sums in Equation~\eqref{eq:MC_efficiency}.  However, these differences in \ac{FAR} threshold have a very small effect on $\langle VT \rangle$:
\GSTLAL and \PYCBC would miss less than $\missedInjectionGstLALPyCBC$ of the recovered injections if changing their \ac{O4a} threshold to $\BBHVTFARThreshold$, while \MBTA would gain less than $\missedInjectionMBTA$.
We consider these differences to be negligible when comparing \ac{O4a} sensitivity estimates to our current results.

\begin{figure*}[htp]
    \centering
    \includegraphics[width=1.0\textwidth]{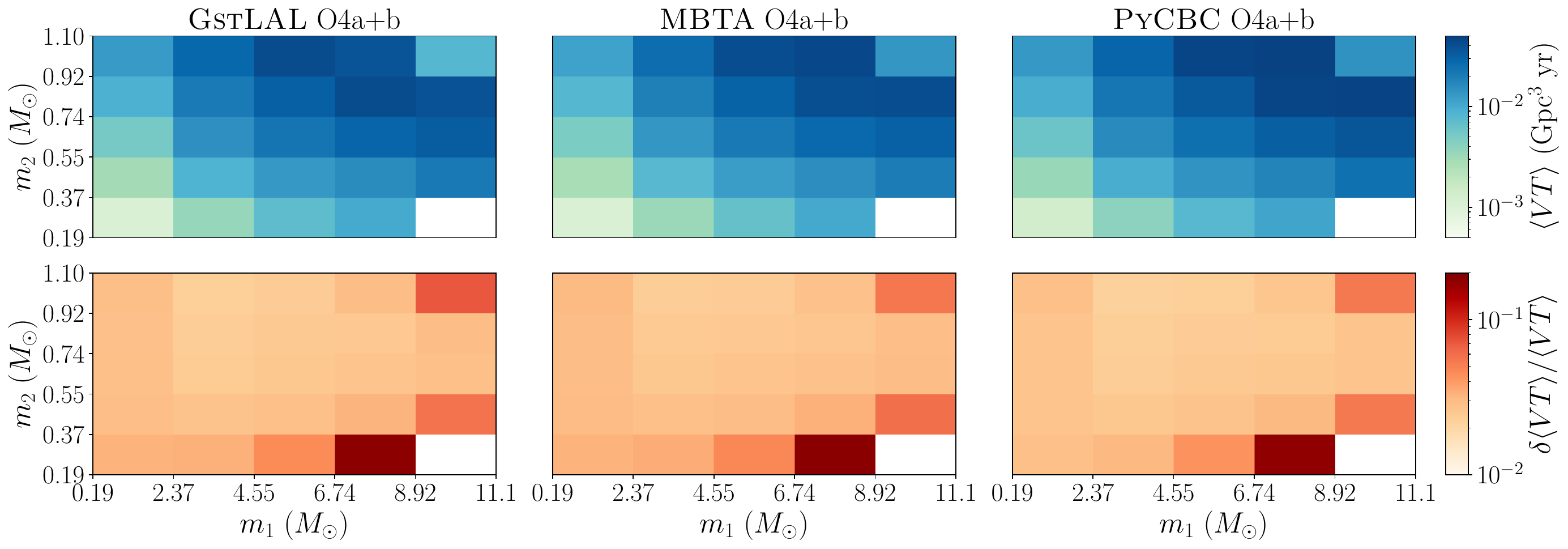}
    \caption{
        Low spin \ac{BBH} population sensitive hypervolume estimates at a \ac{FAR} $\BBHVTFARThreshold$ threshold for each pipeline over the combined \Ofourab runs (top panel), and their corresponding Monte Carlo sampling standard deviations (bottom panel).
        Blank bins correspond to zero efficiency, i.e.\ no injections found by the pipelines with significance above detection threshold.
        }
    \label{fig:o4ab_m1m2}
\end{figure*}

We also compute the sensitivity as a function of component masses $\masscomponent$ for low spins, shown in Figure~\ref{fig:o4ab_m1m2}. 
As expected, search sensitivity appears mainly determined by the binary chirp mass; the highest sensitivity is obtained for the more massive bins, except for the top right (highest total mass) bin which lies beyond the space covered by the search template banks.
Sensitivities are broadly similar between different pipelines. 
While the fractional uncertainty in $\langle VT \rangle$, shown in the bottom panel, is generally small, excess uncertainty in one bin $(\massone,\masstwo)=(\highvariancemOnemin\text{--}\highvariancemOnemax,\highvariancemTwomin\text{--}\highvariancemTwomax)\,\Msun$ is caused by our population model constraint on mass ratio $q\geq 0.05$: this bin then lies at the edge of the population, receiving very few Monte Carlo samples. (The blank bin at lower right lies completely outside our population model and hence cannot be constrained.)

\begin{figure}[h]
    \centering
        \includegraphics[width=1.0\columnwidth]{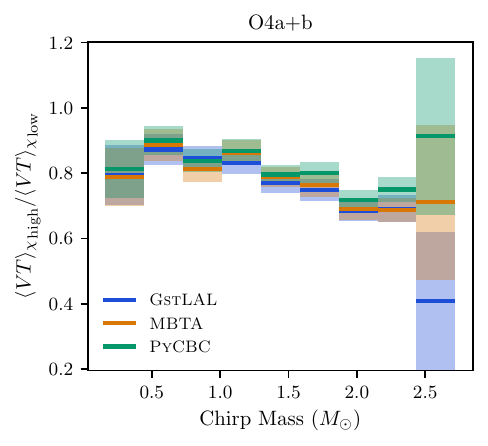}
    \caption{
        Ratio of sensitivities between high- and low-spin populations of \ac{BBH} in chirp mass bins, for the combined \Ofourab runs.
        Uncertainties are propagated from the $\langle VT \rangle$ uncertainty for each population.
        }
    \label{fig:o4ab_ratio_spins}
\end{figure}

Moving to the high-spin case, $\MaxLowSpinMagVTpop\leq\chi_{1,2}\leq\MaxHighSpinMagVTpop$, we find minor changes in the sensitive hypervolume compared to low spins, summarized in Figure~\ref{fig:o4ab_ratio_spins}.
The sensitivity for \ac{O4b} is reduced by $\sim\ratioHightoLowSpinsRunB$, while the combined \Ofourab estimates drop by $\sim\ratioHightoLowSpinsRunAB$.  A weak trend towards lower relative sensitivity for high-spin \acp{BBH} at higher chirp mass is visible, except in the lowest mass bin; note that search template banks cover a reduced range of effective inspiral spin $\chieff$ at the lowest binary masses ($\spincomponent \leq \MaxSpinLowMass$ for $m_{1,2}^{\rm det} \leq \SSMLowHighMassGap\,\Msun$). 
The high-spin population model implies larger statistical uncertainties: reweighting to a uniform distribution in spin magnitudes from the injected distribution (see Section~\ref{sec:3.2-injections}) reduces the number of effective samples, increasing the variance of the Monte Carlo sums by up to a factor of $\increaseVarianceHighSpins$ which translates to a greater uncertainty in $\langle VT \rangle$.

%% file: source__3.2-SSM_BNS.tex
\subsection{Sub-solar mass BNS systems}\label{sec:bns_vt}

Following~\citet{LIGOScientific:2026wxz}, we also report sensitivity estimates for \ac{BNS} coalescences with at least one \ac{SSM} component, using injection waveforms that include tidal effects: this aims to quantify how recovery of such signals is affected by the absence of tidal effects in the search template waveforms. 
We consider \ac{BNS} injections recovered by \MBTA and \PYCBC; in contrast to the \ac{BBH} injection analyses, the \GSTLAL pipeline did not analyze simulated \ac{BNS} signals, in order to limit its total computational cost.

As in~\cite{LIGOScientific:2026wxz}, we extend the sensitivity estimates for \acp{NS} with masses down to $\MinSecondaryMassNS$, in contrast to estimates for \ac{GWTC} releases such as~\cite{2026ApJ..1004L..22A} where the minimum mass considered is $1\,\Msun$. We present the sensitivity for \ac{BNS} systems at several combinations of component masses, considered as representing two different astrophysical scenarios.
\begin{itemize}
    \item Double \ac{SSM} systems: these have both component masses below $1\, \Msun$.
    We show selected points at $\masscomponent = \{0.6,\,0.8,\,1.0\}\, \Msun$.
    \item Single \ac{SSM} systems: for these the secondary is the only \ac{SSM} component, with the primary above $1\, \Msun$. 
    We show estimates at $\massone$ equal to $\{1.4,\,1.8,\,2.2\}\, \Msun$, while $\masstwo$ equal to $\{0.6,\,0.8,\,1.0\}\, \Msun$.
\end{itemize}
To evaluate search sensitivity for each of these combinations, we consider a population of \ac{BNS} systems following a log-normal distribution in each component mass, centered at each $\masscomponent$ point, with a standard deviation of $\LogNormSigmaBNSVT$.

\Ac{NS} spin magnitudes, in contrast to those of \acp{BH}, are not limited to a maximum of $1$; instead, they are constrained to the mass-shedding limit, balancing gravitational force with rotational acceleration.
The most rapidly spinning \ac{NS} observed within a \ac{BNS} system, \FastestBNSName, rotates with a period of $\FastestBNSPeriod$ \citep{2018ApJ...854L..22S}.  Following the dimensionless spin magnitude definition
\begin{equation}
    \label{eq:chi_from_j}
    \chi = \frac{c\ J}{G\ M^2} \quad ,
\end{equation}
and assuming a $1.4\,\Msun$ object and the \GPPVADDtwo \ac{EOS},%
\footnote{In contrast to other authors such as~\citet{Brown:2012qf}, we opt for a moment of inertia corresponding to a solid sphere $I=2/5\,M\,R^2$ dependent on the \ac{EOS}, instead of a constant $I=J/\Omega=\BrownMomentInertia$.} we find a spin magnitude $\chi \approx \FastestBNSSpinMag$.
However, if we instead consider this object a sub-solar \ac{NS} of $0.5\,\Msun$ then its spin would correspond to $\chi \approx \FastestSsmBNSSpinMag$, based on keeping the period fixed while decreasing the mass.
Going further, higher rotational frequencies have been found for \acp{NS} in binary systems with other types of object~\citep[e.g.,][]{Hessels:2006ze}; thus, given the high uncertainties affecting possible \ac{SSM} \ac{BNS} we allow for rapidly-spinning components, comparable to millisecond pulsars~\citep{Bildsten:1997vw}. 
Imposing an upper limit on the \ac{NS} spin magnitude of $\maxspinBNS$, consistent with~\cite{2026ApJ..1005L..51A,LIGOScientific:2026wxz}, our sensitivity estimates consider a uniform distribution in spin magnitude constrained to $\chi_{1,2} \le \maxspinBNS$, and isotropic spin orientations.

\begin{figure}[h]
    \centering
        \includegraphics[width=1.0\columnwidth]{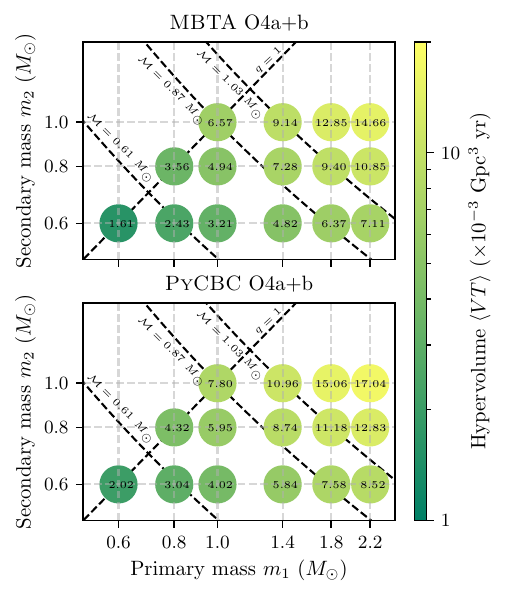}
    \caption{
        Sensitive hypervolume estimates for a \ac{SSM} \ac{BNS} population at a \ac{FAR} $\BNSVTFARThreshold$ threshold. We report the sensitivities of the \MBTA and \PYCBC pipelines over the combined \Ofourab runs evaluated at points in the component mass space.
        }
    \label{fig:o4ab_bns_vt}
\end{figure}

As in the previous section, we choose a reference \ac{FAR} threshold of $\BNSVTFARThreshold$ to present sensitivity estimates for each pipeline.
Figure~\ref{fig:o4ab_bns_vt} shows the sensitive hypervolume computed at the proposed points in the $\masscomponent$ space.
We observe an increase between a factor of $\increaseLowBoundBNSvt$ and $\increaseHighBoundBNSvt$ when comparing the combined estimate to previous \ac{O4a} results~\citep{LIGOScientific:2026wxz}.
Uncertainties propagated from the Monte Carlo sum variance range between $\lowUncertBNSVT$ and $\highUncertBNSVT$ for the most and least massive $\masscomponent$ points, respectively.

Comparing \ac{SSM} \ac{BBH} and \ac{BNS} with the same population mass and spin distributions, we find that sensitivity to \acp{BNS} can be slightly lower, ranging over $\sim\!(\fractionBBHtoBNS\text{--}1)$ times the \ac{BBH} sensitivity.
Loss in sensitivity for \ac{BNS} is primarily caused by the mismatch between the templates (see Section~\ref{sec:2-search}) and injected waveforms including tidal effects.  In the case of super-solar \acp{NS} such mismatch is expected to be subdominant~\citep{Williams:2022vct} and restricted to a small portion of the parameter space, which would appear with a fitting factor below the designed minimal match of the bank~\citep{Piccari:2025itn}.
However, given we employ the stiffest \ac{EOS} and consider masses in the sub-solar regime, this effect is magnified and can lead to significant sensitivity loss due to template bank mismatch~\citep{Bandopadhyay:2022tbi}.
Additionally, waveforms including tidal effects are not calibrated in this low mass range: thus, tidal deformations in \ac{SSM} \acp{NS} are extrapolated into the sub-solar regime, which may introduce systematic uncertainty in our sensitivity estimates due to waveform errors.

%% file: source__4-rates.tex
\section{Merger rates}\label{sec:4-rates}

The $\langle VT \rangle$ sensitivity estimates obtained in the previous section, coupled with the non-detection search results of Section~\ref{sec:results}, can be used to derive upper limits on the merger rate of possible populations of low-spin \ac{SSM} \acp{BBH} or \acp{BNS}. Hereafter, the rates and constraint results quoted for \ac{SSM} \acp{BBH} are for the low-spin population. Assuming the physical independence of different mergers, we treat them as an inhomogeneous Poisson process, with an intensity varying over binary component masses and spins and over redshift. We make the further assumption, reasonable given the maximum redshift $z_{\rm max} \sim \MaxRedshiftInjOfourb$ probed by our searches, that the merger rate is constant over comoving volume and source-frame time, notating the total rate density of \acp{BBH} over the \ac{SSM} parameter space as $\mathcal{R}$.\footnote{A \ac{SSM} \ac{BNS} population is described similarly, but has its own independent rate parameter.} 

A model of the \ac{SSM} \ac{BBH} population $\mathbf{\Lambda}$ is then described by a normalized distribution of mergers over intrinsic parameters (component masses and spins), $p(\boldsymbol{\theta}|\mathbf{\Lambda})$.  If identified by a search pipeline, a merger signal is assigned a ranking parameter $x$ such that louder signals correspond to higher $x$ values; here, we may identify $x$ with the \acl{IFAR}.  The expected number of mergers identified by a search above some threshold value of $x$ is then given by 
\begin{equation}
 \mu(x,\mathbf{\Lambda}) = \mathcal{R} VT(x,\mathbf{\Lambda}) \,, 
\end{equation}
where we explicitly indicate the dependence of $VT$ on the threshold and on the model $\mathbf{\Lambda}$, cf.\ Equation~\eqref{eq:VT_from_xi}.  Search candidates due to astrophysical mergers, which are assigned a ranking above a threshold $x$, can then be treated as a Poisson process with parameter $\mu(x,\mathbf{\Lambda})$.

We then obtain upper limits on $\mathcal{R}$ by the (frequentist) loudest event statistic method~\citep{2004CQGra..21S1775B}: the value of $\mu$ is bounded above by the fact that there were no candidates identified with higher ranking than the loudest event recorded by a search, $\hat{x}$.  Imposing the ranking threshold $\hat{x}$, we have for a $p\%$ confidence upper limit $\mathcal{R}_p$,
\begin{equation} \label{eq:rate_limit}
  1 - \frac{p}{100} = \exp(-\mu_p(\hat{x},\mathbf{\Lambda})) \implies 
  \mathcal{R}_p = -\frac{\ln(1 - p/100)}{VT(\hat{x},\mathbf{\Lambda})}\,,
\end{equation} 
yielding a standard $90\%$ confidence upper limit $\mathcal{R}_{90}\simeq 2.3 \langle VT(\hat{x},\mathbf{\Lambda})\rangle^{-1}$.  We replace the sensitivity $VT$ by its estimated value $\langle VT \rangle$ assuming its statistical uncertainty is negligible.

\begin{figure*}[tbp]
    \centering
    \includegraphics[width=1.0\textwidth]{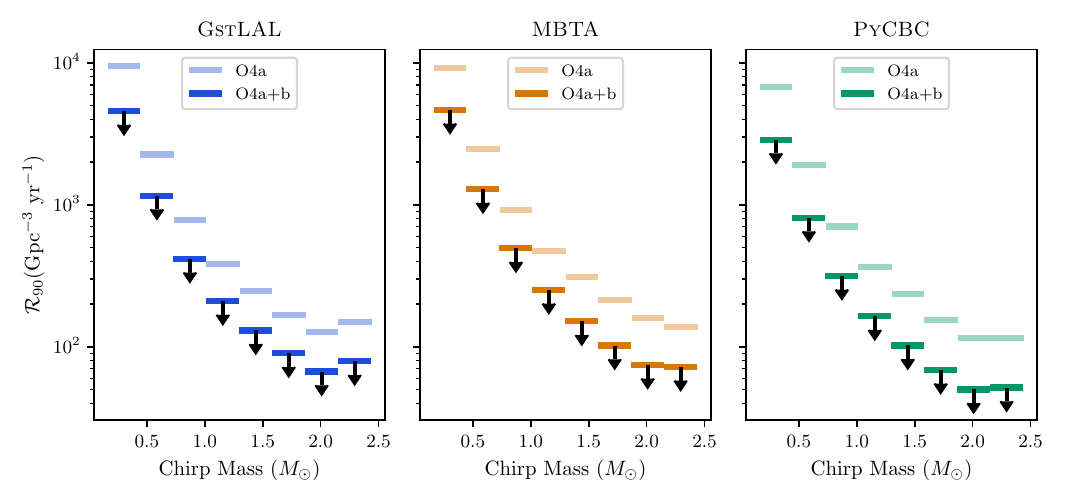}
    \caption{
        $90\%$ confidence upper limit $\mathcal{R}_{90}$ on the merger rate for a \ac{BBH} population, inferred for \ac{O4a} and the combined \Ofourab runs, displayed for each pipeline as a function of chirp mass. Arrows point in the direction of the possible merger rate not constrained by the \Ofourab search.
        }
    \label{fig:o4a_o4ab_rates}
\end{figure*}
Adopting the loudest event formalism as above, we obtain upper limits on the merger rate for each pipeline considering the combined \Ofourab runs.
We estimate the sensitive hypervolume (Section~\ref{sec:3-search_sensitivity}) at the \ac{FAR} threshold determined by the loudest event recorded by each pipeline over both the \ac{O4a} and \ac{O4b} runs, and follow Equation~\eqref{eq:rate_limit} to derive $90\%$ confidence upper limits.
\PYCBC found a louder event during \ac{O4a}~\citep{LIGOScientific:2026wxz} compared to the most significant event in \ac{O4b} (see Section~\ref{sec:results}) while \GSTLAL and \MBTA found the loudest events in \ac{O4b}. 
Therefore, we use the following \ac{FAR} thresholds: $\GSTLALLoudestTriggerFAR$ for \GSTLAL, $\MBTALoudestTriggerFAR$ for \MBTA, and $\PYCBCLoudestTriggerFARCombined$ for \PYCBC. 
This is in contrast to Sections~\ref{sec:bbh_vt} and \ref{sec:bns_vt}, where sensitivity estimates adopted a constant $\BBHVTFARThreshold$ \ac{FAR} threshold.

Figure~\ref{fig:o4a_o4ab_rates} shows the upper limit on the \ac{SSM} \ac{BBH} merger rate at the $90\%$ confidence level for \GSTLAL, \MBTA, and \PYCBC as a function of the chirp mass, \Mc, and the relative improvement between the different observing periods.
Since a high-spin population of \ac{SSM} \ac{BBH} shows only minor variation in $\langle VT \rangle$ with respect to a low-spin population (see Section~\ref{sec:bbh_vt}), we show upper limits only for the low-spin case, following previous studies~\citep{LIGOScientific:2026wxz}.
As expected from Equation~\eqref{eq:rate_limit},
our rate limits are more constraining by a factor comparable to the increases in $\langle VT \rangle$ from \ac{O4a} to \Ofourab.
The lowest upper limits $\mathcal{R}_{90}$ are $\MostConstrainedRateMcGSTLAL$, $\MostConstrainedRateMcMBTA$, and $\MostConstrainedRateMcPYCBC$ for \GSTLAL, \MBTA, and \PYCBC respectively.

\begin{figure*}[tbp]
    \centering
    \includegraphics[width=1.0\textwidth]{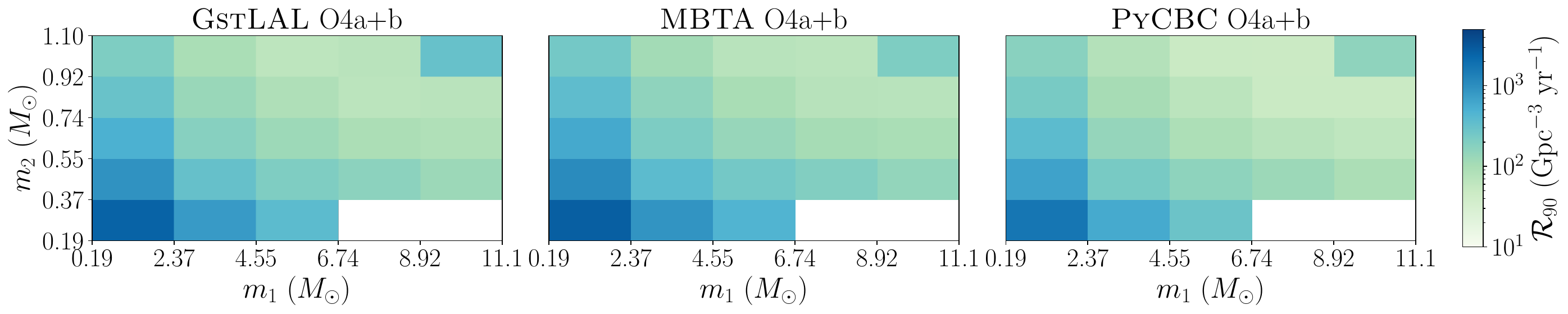}
    \caption{
        $90\%$ confidence upper limit $\mathcal{R}_{90}$ on the merger rate as a function of $\masscomponent$ for the combined \Ofourab runs, displayed for each pipeline.
        }
    \label{fig:o4ab_rates_m1m2}
\end{figure*}

We also report upper limits on the merger rate as a function of $\masscomponent$, following the binning 
described in Section~\ref{sec:bbh_vt}, as shown in Figure~\ref{fig:o4ab_rates_m1m2}.
We find the most constraining limit $\mathcal{R}_{90}$ to be $\MostConstrainedRateCompmassGSTLAL$, $\MostConstrainedRateCompmassMBTA$, and $\MostConstrainedRateCompmassPYCBC$ for \GSTLAL, \MBTA, and \PYCBC respectively.
Blank bins correspond to two cases: the $(\massone,\masstwo)=(6.74{-}8.92,0.19{-}0.37)~\Msun$ bin, which is found to have an excessive variance (see Figure~\ref{fig:o4ab_m1m2}), and the $(\massone,\masstwo)=(8.92{-}11.10,0.19{-}0.37)~\Msun$ bin, which corresponds to zero efficiency.

\begin{figure}[h]
    \centering
        \includegraphics[width=0.9\columnwidth]{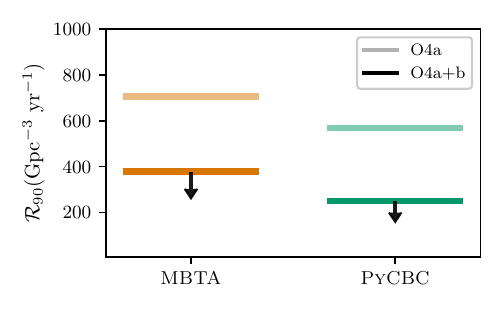}
    \caption{
        $90\%$ confidence level upper limit on a broad population of \ac{SSM} \acp{BNS} reported by \PYCBC and \MBTA during \ac{O4a} and combined \Ofourab runs.  Arrows point in the direction of the possible merger rate not constrained by the \Ofourab search.
        }
    \label{fig:o4a_o4ab_bns_rates}
\end{figure}

Additionally, we report the upper limit on the \ac{SSM} \ac{BNS} merger rate at the $90\%$ confidence level based on the \MBTA and \PYCBC results, assuming a simple fixed population model as in~\citet{2026ApJ..1005L..51A} (\textsc{Simple Uniform BNS}). 
This consists of a uniform mass distribution between $\MinPrimaryMassNS$ and $\MaxPrimaryMassNS$ for $\massone$, and a uniform mass distribution between $\MinSecondaryMassNS$ and $\MaxSecondaryMassNSVT$ for $\masstwo$ with $\masstwo \leq \massone$, isotropically distributed spins with uniform spin magnitudes below $\maxspinBNS$, and a merger rate uniform in comoving volume up to $\redshift_{\rm max}=\MaxRedshiftInjNS$.
Following the same loudest event formalism we find the upper limit at the 90\% confidence level to be $\SSMBNSRateOFourAB$.
In Figure~\ref{fig:o4a_o4ab_bns_rates} we show the relative improvement from \ac{O4a} and combined runs in the estimation of the upper limit.
Our inferred \ac{SSM} \ac{BNS} merger rate limit lies at the upper limit of the rate interval estimated in~\citep{2026ApJ..1005L..51A} (\GWTCfourBNSRate) for \ac{BNS} with masses above $1~\Msun$, assuming the \textsc{Simple Uniform BNS} distribution.
While these rates are not strictly comparable due to the different \ac{NS} mass ranges considered, we cannot rule out that the rate of \ac{SSM} \ac{BNS} is close to that of super-solar mass \ac{BNS}.

%% file: source__5-dm_constraints.tex
\section{Dark Matter Constraints}\label{sub:5-dm_constraints}

The upper limits on merger rates obtained in Section~\ref{sec:4-rates} are the main results derived from our \ac{SSM} search through \ac{O4a} and \ac{O4b}, being the most restrictive to date.
These limits can be used to constrain any model that predicts a non-negligible rate of binary mergers within the parameter space analysed by the search.
In particular, we use the rates and sensitivities of low-spin \ac{SSM} \acp{BBH} for the constraints presented in the following sections. 
Following~\citet{LIGOScientific:2026wxz}, in the next subsections we update the constraints on two possible sources of \ac{BBH} mergers: \aclp{PBH} in Section~\ref{sec:5.1-PBH} and dissipative dark matter black holes in Section~\ref{sec:5.2-DDM}.

\input{source__5.1-PBH.tex}

\input{source__5.2-DDM.tex}

%% file: source__5.1-PBH.tex
\subsection{Primordial Black Holes}\label{sec:5.1-PBH}

\Acp{PBH} are natural dark matter candidates that could give rise to \ac{BBH} mergers in the \ac{SSM} range. Many models have been proposed~\citep{Carr:2023tpt,LISACosmologyWorkingGroup:2023njw} for which the \ac{PBH} abundance is characterized by the fraction of the dark matter cosmological density in the form of \acp{PBH}, $f_\mathrm{PBH} = \rho_\mathrm{PBH}/\rho_\mathrm{DM}$, where $\rho_\mathrm{PBH}$ and $\rho_\mathrm{DM}$ are the present-day mass densities of \acp{PBH} and dark matter, respectively. Models also differ in the shape of the \ac{PBH} mass distribution, which depends in detail on specific formation mechanisms. Furthermore, \ac{PBH} models producing \ac{GW} signals today involve environments in which binaries can form and merge.

Possible constraints on $f_\mathrm{PBH}$, based on our \ac{BBH} merger rate upper limits, are thus particularly dependent on model assumptions. 
First of all, merger rates depend strongly on the initial (extended) mass distributions, which determine the relative abundance of mass combinations in the binaries and the properties of the PBH clusters. Secondly, the merger rate predictions are strongly influenced by complex phenomena of the formation environment, such as many-body interactions and non-linear gravitational effects, which are difficult to model accurately.  Finally, specifically for \ac{SSM} \acp{BH}, the merger times may be so much longer than the age of the universe that the probability of detecting a single merger in a given \ac{LVK} observation run is negligible. 
\Ac{PBH} binaries are commonly assumed to form in two different scenarios, that we refer as \acl{EB} and \acl{LB}. \Ac{EB} bind shortly after \ac{PBH} formation as pairs of \acp{PBH} decouple from the Universe's expansion and form eccentric binaries through tidal interactions with nearby \acp{PBH}~\citep{Nakamura:1997sm}. Consequently, merger rates of \ac{EB} are strongly influenced by the initial \ac{PBH} clustering and mass distribution~\citep{Carr:2026hot}.
In contrast, \ac{LB} form in dense halos through dynamical interactions, with merger rates shaped by local density and halo properties~\citep{Siles:2024yym}.

To illustrate how the upper limit of merger rates can be mapped to constraints on an underlying astrophysical population of \acp{PBH}, we model the merger rate of \ac{EB} as in~\citet{2023MNRAS.524.5984L}, using the analytical approximations developed in~\citet{Hutsi:2020sol,Vaskonen:2019jpv,Chen:2018czv,Ali-Haimoud:2017rtz} and parametrized with $N$-body simulations via~\citet{Raidal:2018bbj}
\begin{align}
	&\frac{\mathrm{d}^2\mathcal{R}^\mathrm{PBH}_\mathrm{EB}}{\mathrm{d} \ln m_A \mathrm{d} \ln m_B} = 1.6 \times 10^6 \, \mathrm{Gpc}^{-3} \, \mathrm{yr}^{-1} \times f_\mathrm{sup} \,f_\mathrm{PBH}^{53/37}   \nonumber \\
	& \times f(\ln m_A)f(\ln m_B)\,\left( \frac{m_A+m_B}{\Msun} \right)^{-32/37} \left[ \frac{m_A m_B}{(m_A+m_B)^2}\right]^{-34/37},  \label{eq:Rate_EB}
\end{align}
where $m_A$ and $m_B$ denote the component masses of the \ac{PBH} binary, without imposing any ordering between them, in contrast to the convention $\massone \geq \masstwo$ used elsewhere. The normalized \ac{PBH} mass function $f(\ln m) = 1/\rho_{\rm DM} \cdot d\rho_\mathrm{PBH}/d\ln m$ is defined as the differential PBH energy density with respect to $\ln m$, normalised to the total energy density in DM.  The suppression factor $f_\mathrm{sup}$ accounts for binary disruption due to gravitational interactions with other \acp{PBH} or surrounding matter. The merger-rate expression of Equation~\eqref{eq:Rate_EB}, while widely used in the literature, has only been validated by $N$-body simulations in the case of monochromatic and narrow peaked mass distributions.
The simulations of~\citet{Raidal:2018bbj} consider $\sim\mathcal{O}(10^2)$ equal-mass $\sim 30\,\Msun$ \acp{PBH} with $f_\mathrm{PBH} \lesssim 0.1$, which differs significantly from the lower masses and larger \ac{PBH} abundances explored here. Further developments in \ac{EB} formation may substantially change the suppression factor in the case of extended mass functions; in this study we adopt $f_\mathrm{sup} = 2.3 \times 10^{-3} f_\mathrm{PBH}^{-0.65}$ to remain consistent with~\citet{2023MNRAS.524.5984L}, which represents a high-suppression scenario~\citep{Hutsi:2020sol}, and thus yields conservative constraints, in the monochromatic mass case.

Predicting the merger rate of \ac{LB} involves the modeling of complex dynamical interactions. Following~\citet{Clesse:2020ghq,Phukon:2021cus}, we assume
\begin{align}
	\frac{\mathrm{d}^2\mathcal{R}^\mathrm{PBH}_\mathrm{LB}}{\mathrm{d} \ln m_A \mathrm{d} \ln m_B} = R_\mathrm{clust} f_\mathrm{PBH} f(\ln m_A) f(\ln m_B) \frac{(m_A + m_B)^{10/7}}{(m_A m_B)^{5/7}},  \label{eq:Rate_LB}
\end{align}
where $R_\mathrm{clust}$ is a scaling factor that depends on the \ac{PBH} clustering properties, including their velocity distribution. While there is general agreement on the mass dependence of the \ac{LB} merger rate~\citep[e.g.,][]{Clesse:2016vqa}, the overall amplitude remains highly uncertain due to its strong sensitivity to the underlying clustering scenario. As in~\citet{Clesse:2020ghq,Phukon:2021cus}, and to remain consistent with~\citet{2023MNRAS.524.5984L}, we assume $R_\mathrm{clust} = 420 \, \mathrm{Gpc}^{-3} \, \mathrm{yr}^{-1}$; this overall factor can  be easily rescaled for different assumptions or clustering models.

\begin{figure}
    \centering
    \includegraphics[width=1\columnwidth]{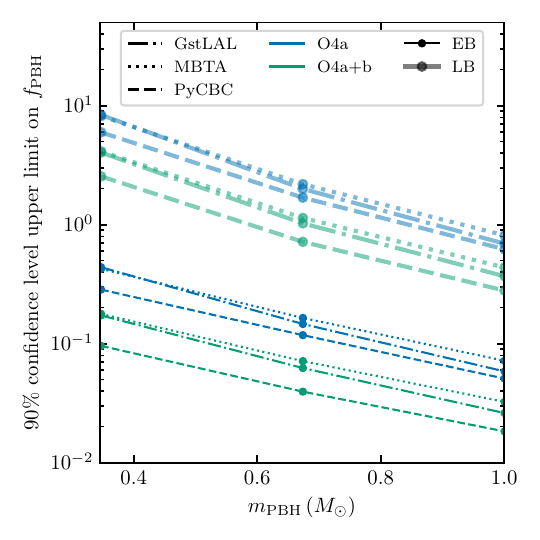}
    \caption{
    Upper limits on the \ac{PBH} dark matter fraction $f_{\text{PBH}}$ as a function of \ac{PBH} mass $m_{\text{PBH}}$.
    Constraints are derived using the merger rate models given by Equation~\eqref{eq:Rate_EB} for \acl{EB}, and Equation~\eqref{eq:Rate_LB} for \acl{LB}, assuming monochromatic mass functions.
    Results are shown at the 90\% confidence level using the sensitive hypervolumes $\langle VT \rangle$ calculated for \GSTLAL, \MBTA, and \PYCBC for the combined O4a+O4b runs. Previous \ac{O4a} results are also shown for comparison.
    }
    \label{fig:fPBH_O4b_withLB}
\end{figure}

We use the theoretical merger rate models of Equations~(\ref{eq:Rate_EB},~\ref{eq:Rate_LB}) to set upper limits on $f_\mathrm{PBH}$, assuming a monochromatic \ac{PBH} mass distribution, i.e., $f(\ln m) = \delta(\ln m - \ln m_\mathrm{PBH})$.
The limits are obtained by identifying the value of $f_\mathrm{PBH}$ for which the theoretical rates match the 90\% confidence upper bounds on merger rates derived using Equation~\eqref{eq:rate_limit}.
Figure~\ref{fig:fPBH_O4b_withLB} shows the resulting upper limits for both \ac{EB} and \ac{LB} for the \GSTLAL, \MBTA, and \PYCBC pipelines searching the combined  \Ofourab run, as well as the \ac{O4a} results for comparison. The upper limits obtained in the combined \Ofourab run are approximately half of the \ac{O4a} ones, reflecting the comparable duration of the \ac{O4a} and \ac{O4b} observing periods.
The combined \Ofourab runs provide the strongest \ac{LVK} limits to date on the \ac{PBH} fraction, $f_\mathrm{PBH} \leq \EBfPBHMaxMPBHOfourAB$ at \MaxMPBH and $f_\mathrm{PBH} \leq \EBfPBHMinMPBHOfourAB$ at \MinMPBH.
These direct search constraints are complementary to the bounds derived from the non-observation of a stochastic \ac{GW} background from inspiralling \ac{SSM} \ac{PBH} binaries at high redshift~\citep{Boybeyi:2024mhp,LIGOScientific:2025kry}.

While the constraints presented above assume a monochromatic \ac{PBH} mass function, this is not realistic in most early formation scenarios~\citep{Carr:2023tpt}.
\acp{PBH} are generally expected to form with an extended and non-trivial mass distribution~\citep{LISACosmologyWorkingGroup:2023njw}.
To remain agnostic about the specific shape of the mass function, we instead constrain the quantity
\begin{align}
	F_\mathrm{PBH} & = \frac{f_\mathrm{sup}}{2.3\times 10^{-3}} f_\mathrm{PBH}^{53/37} f(\ln m_A) f(\ln m_B) \nonumber \\
	& = f_\mathrm{PBH}^{0.78} f(\ln m_A) f(\ln m_B) ,  \label{eq:FPBH_def}
\end{align}
which encapsulates the dependence of the merger rate of \ac{EB} on both the \ac{PBH} abundance $f_\mathrm{PBH}$, and the underlying mass distribution $f(\ln m)$.
The definition of $F_\mathrm{PBH}$ explicitly incorporates the assumption $f_\mathrm{sup} = 2.3 \times 10^{-3} f_\mathrm{PBH}^{-0.65}$ adopted throughout this study.

Limits on $F_\mathrm{PBH}$ are determined by finding the value for which the theoretical merger rate for \ac{EB} matches the 90\% confidence level upper bounds derived from Equation~\eqref{eq:rate_limit}.
We constrain $F_\mathrm{PBH}$ only for \ac{EB}, since its definition in Equation~\eqref{eq:FPBH_def} is specific to this scenario, and merger rate models for \ac{LB} are more uncertain and less constraining.
We adopt the binning in (\massone, \masstwo) from Section~\ref{sec:4-rates}, and write the integral in Equation~\eqref{eq:Rate_EB} within each bin as
\begin{align}
	\mathcal{R}^\mathrm{PBH}_\mathrm{EB} = & 3.68 \times 10^3 \, \mathrm{Gpc}^{-3} \, \mathrm{yr}^{-1} \overline{F}_\mathrm{PBH} \times  \left( \frac{\overline{m}_1 + \overline{m}_2}{\Msun} \right)^{-32/37} \nonumber \\
	& \times \left[ \frac{\overline{m}_1 \overline{m}_2}{(\overline{m}_1 + \overline{m}_2)^2}\right]^{-34/37}  \ln\left( \frac{m_1^\mathrm{max}}{m_1^\mathrm{min}}\right) \ln\left( \frac{m_2^\mathrm{max}}{m_2^\mathrm{min}}\right) ,  \label{eq:Rate_EB_FPBH}
\end{align}
where $\masscomponent^\mathrm{min}$ and $\masscomponent^\mathrm{max}$ denote the bin edges for the component masses, $\overline{m}_{1,2} = (\masscomponent^\mathrm{min} + \masscomponent^\mathrm{max})/2$ is the central value, and $\overline{F}_\mathrm{PBH}$ denotes the effective bin-averaged value of $F_\mathrm{PBH}$, defined implicitly such that Equation~\eqref{eq:Rate_EB_FPBH} reproduces the exact integral of the merger rate over the bin.

\begin{figure}
    \centering
\includegraphics[width=1\columnwidth]{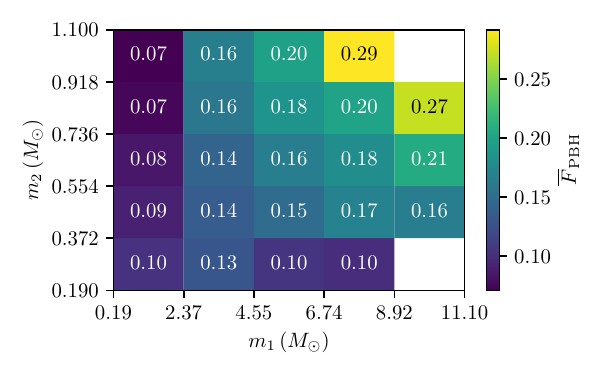}
    \caption{Upper limits on $\overline{F}_\mathrm{PBH}$, Eq.~\eqref{eq:FPBH_def}, as a function of the \ac{PBH} binary component masses \massone\ and \masstwo.
	Constraints are derived using the early binary merger rate model of Equation~\eqref{eq:Rate_EB_FPBH}.
    We show the median 90\% upper limit across \GSTLAL, \MBTA, and \PYCBC calculated using their sensitive hypervolumes $\langle VT \rangle$ in the combined \ac{O4a} and \ac{O4b} runs.}   \label{fig:FPBH_m1m2_O4ab_EB}
\end{figure}

Figure~\ref{fig:FPBH_m1m2_O4ab_EB} 
shows the median 90\% upper limit on $\overline{F}_\mathrm{PBH}$ across all search pipelines, based on the results from the combined runs \Ofourab.
We find that $\overline{F}_\mathrm{PBH}$ is constrained to be less than unity across the entire range of (\massone, \masstwo) explored in this study, with upper limits a factor two below those obtained from \ac{O4a}.

%% file: source__5.2-DDM.tex
\subsection{Dissipative Dark Matter}\label{sec:5.2-DDM}

The method to constrain the fraction of dissipative dark matter in the binary dark matter budget is nearly identical to that used in the analyses of~\citet{LIGOScientific:2026wxz}, but with a key change in that we now report upper limits on $f_\text{DBH}$ conditional on $M_\text{min}$, and the complementary bound on $M_\text{min}$ conditional on $f_\text{DBH}$, rather than a two-dimensional credible region.
Each conditional bound therefore depends only on the posterior at the given value of $M_\text{min}$ (or $f_\text{DBH}$), not on how the posterior behaves elsewhere in the parameter space.
In this section, we reiterate the methods and update the constraints with the search pipeline sensitivity, which combines results from the \ac{SSM} search in \ac{O4a} data and \ac{O4b} data.  

Dissipative dark matter is a class of models which posit that dark matter may have complex particle interactions similar to those of the Standard Model, including dissipative processes.
Whenever there is a fermion in the model, dark matter has a Chandrasekhar limit analogous to that of the Standard Model, but with a value that depends on the dark matter particle mass(es)~\citep{Mohapatra:1999ih, DAmico:2017lqj, Shandera:2018xkn, Choquette:2018lvq, Chang:2018bgx, Bramante:2023ddr}.
In dissipative scenarios, particle interactions allow energy to be transported away from gravitationally bound systems such as dark matter halos.
These dissipative interactions become more efficient in denser regions, and some dense regions may dissipate energy efficiently enough for gravity to dominate, leading to the formation of ultra-compact objects made of dark matter.
When those are sufficiently massive, above the dark Chandrasekhar limit, they are \aclp{DBH}.
Therefore, \ac{GW} signals from \ac{DBH} binaries provide a way to probe the microphysics of dissipative dark matter even if it is not coupled to the Standard Model beyond gravity, or if the coupling is too weak to be detectable in the near future.

Although the fragmentation scale of the dark gas depends on the chemistry of the dark sector, the dynamical processes that lead to the formation of compact objects are analogous to those in star formation.
While simulation of such small-scale phenomena in novel dark matter scenarios remains challenging, it is reasonable to expect that the initial mass function and binary fraction of \acp{DBH} are qualitatively similar to that of baryonic objects.
For the particular worked example of atomic dark matter~\citep{Ackerman:2008kmp, Kaplan:2011yj, Cyr-Racine:2012tfp}, where cooling and compact object formation has been extensively studied~\citep{Rosenberg:2017qia,Ryan:2021dis, Gurian:2021qhk, Ryan:2021tgw}, there is a close parallel between \ac{DBH} formation and Population III star formation that further supports this statement.
Thus, we use a truncated power-law mass model characterized by a minimum mass $M_\text{min}$, power $-b$, and ratio of maximum mass to minimum mass $r = M_\text{max}/M_\text{min}$, and assume a binary fraction of $f_\text{binary} = \DBHfbinary$~\citep{Shandera:2018xkn} in our analysis.
The total number of \acp{DBH} in this population is set by $f_\mathrm{DBH} \equiv \rho_\text{DBH}/\rho_\text{DM}$ where $\rho_\text{DBH}$ and $\rho_\text{DM}$ are the densities of \acp{DBH} and dark matter, respectively.
The rate of \ac{DBH} binaries merging in this population, $\mathcal{R}$, is computed as~\citep{Singh:2020wiq}
\begin{multline}
 \frac{\mathrm{d}\mathcal{R}(\mathcal{M}\mid \, f_{\mathrm{DBH}}, \theta)}{\mathrm{d}\mathcal{M}}
 = \Bigl[ p(\mathcal{M} \mid t_{\rm m}, \theta) p(t_{\rm m} \mid \theta) \Bigr]_{t_{\rm m} = \DBHtmerger} \\
 \times \left(\frac{ f_{\mathrm{DBH}} \, \rho_{\mathrm{DM}} \, f_{\mathrm{binary}}}{\langle M \rangle}\right)\,,
\label{eq:peters_formula}
\end{multline}
where $p(\mathcal{M} \mid t_{\rm m}, \theta)$ is the distribution of chirp mass with a time of merging $t_{\rm m}$, $p(t_{\rm m} \mid \theta)$ is the probability distribution of times-to-merger (these two factors evaluated at $t_{\rm m} = \DBHtmerger$), $\rho_\text{DM} = \DBHrhoDM$~\citep{Planck:2018vyg}, and $\langle M \rangle$ is the average component mass of the \ac{DBH} binaries.
The merger times are calculated using Peters' formula for coalescing binaries~\citep{1963PhRv..131..435P}.

We use a Bayesian analysis to constrain this population of \acp{DBH} by modeling the number of detected binaries with a Poisson distribution in each of \DBHMcBinCount uniformly spaced chirp mass bins spanning $\DBHMcBinLo \leq \Mc \leq \DBHMcBinHi$.
With the non-detection of significant \ac{SSM} candidates, the likelihood in the $i$-th chirp-mass bin is $P(0|\mathcal{R}_i,\langle VT \rangle_i,\Mc) = e^{-\mathcal{R}_i \langle VT \rangle_i}$, where the rate $\mathcal{R}_i$ is determined by integrating Equation~\eqref{eq:peters_formula} over the width of the $i$-th chirp mass bin, and the sensitive hypervolume $VT_i$ is informed by this \ac{SSM} search.
To ensure an overall non-detection result, which will be applicable to each chirp mass bin, we adopt a more stringent \ac{FAR} threshold of \FARThreshVTDBH when estimating the pipeline-specific $\langle VT \rangle_{\rm O4a+b}$ for this analysis.
Additionally, we are discarding \MBTAFirstTriggerSEventName which has a \ac{FAR} below \FARThreshVTDBH due to it being consistent with noise as detailed in Section~\ref{sec:results}.

We use priors consistent with previous analyses~\citep{2022PhRvL.129f1104A,2023MNRAS.524.5984L}: uniform in log-space priors on $M_\text{min}$ over the interval $[10^{-3}, 3.1]\,\Msun$, and $f_\text{DBH}$ over the interval $[10^{-10},1]$, and a log-uniform prior on $r$ over the interval $[2, 1000]$, as well as a uniform in linear-space prior on $b$ over the interval $[-1,2]$.
From Bayes' theorem, the log-posterior is then
\begin{multline}
 \ln p(f_\text{DBH}, M_\text{min}, r,b) = \\ 
 -\sum_i \mathcal{R}_i(f_\text{DBH},M_\text{min},r,b) \langle VT \rangle_i - \ln r + C\,,
\end{multline}
where $C$ is a normalization constant.

\begin{figure}
	\centering
	\includegraphics[width=1\columnwidth]{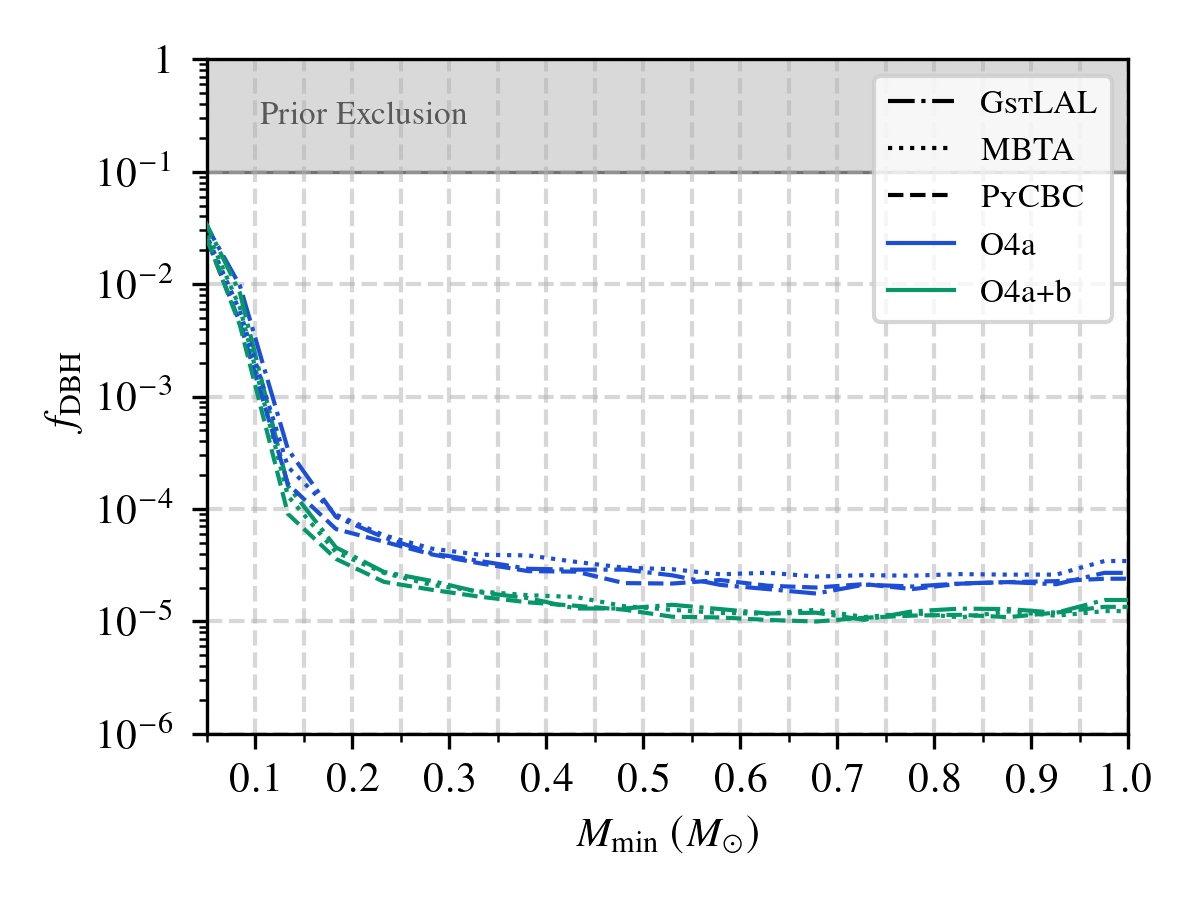}
	\caption{
		Conditional 90\% upper limits on $f_{\rm DBH}$ as a function of $M_{\min}$ for the atomic dark matter model, after marginalizing over the ratio of maximum to minimum mass and the slope of the black hole population.
		The samples are binned in the $f_{\rm DBH}$--$M_{\min}$ plane, and within each $M_{\min}$ bin the posterior is normalized and integrated in $f_{\rm DBH}$ following Equation~\eqref{eq:conditional_upper_limit}; values of $f_{\rm DBH}$ above each curve are excluded.
		The shaded band marks the 90\% quantile of the $f_{\rm DBH}$ prior, shown for reference as the level a limit approaches where the data is uninformative.
		Constraints are shown for the \GSTLAL, \MBTA, and \PYCBC pipelines, derived from $\langle VT\rangle_{\rm O4a}$ and from the combined $\langle VT\rangle_{\rm O4a+b}$, each evaluated with a \ac{FAR} threshold of \FARThreshVTDBH.
		The improvement from $\langle VT\rangle_{\rm O4a}$ to $\langle VT\rangle_{\rm O4a+b}$ is roughly a factor of two across the range where the data best constrains $f_{\rm DBH}$.
		}
	\label{fig:atomic_dark_matter_f_vs_mmin}
\end{figure}

Analyses through the end of \ac{O3b} evaluated the posterior distribution using a grid-based method~\citep{2022PhRvL.129f1104A,2023MNRAS.524.5984L}, while we use the \ac{MCMC} sampler \textsc{emcee}~\citep{Foreman-Mackey:2012any} to explore the parameter space for faster convergence in this analysis.
We marginalize over $r$ and $b$ to constrain $f_\text{DBH}$ and $M_\text{min}$.
The resulting posterior is two-dimensional; we then compute the one-dimensional probability density over $f_\text{DBH}$ conditioned on $M_\text{min}$ and use this to obtain the 90\% upper bound $f_{\text{DBH},90\%}(M_\text{min})$ by solving
\begin{equation}
    \int_{0}^{f_{\text{DBH},90\%}(M_\text{min})} p(f_\text{DBH} \mid M_\text{min})\, \mathrm{d}f_\text{DBH} = 0.9
    \label{eq:conditional_upper_limit}
\end{equation}
for each $M_\text{min}$, where $p(f_\text{DBH} \mid M_\text{min}) = p(f_\text{DBH}, M_\text{min}) / \int p(f_\text{DBH}, M_\text{min})\, \mathrm{d}f_\text{DBH}$.
Both integrals are evaluated as sums over the $f_\text{DBH}$ bins of the two-dimensional sample histogram.
The $\langle VT \rangle$ values are updated relative to~\citet{LIGOScientific:2026wxz} by incorporating $\langle VT \rangle_{\rm O4a+b}$, and the $\langle VT \rangle_{\rm O4a}$ constraints quoted here are recomputed from the corresponding posterior samples using this method, so they differ from the values reported there.
Figure~\ref{fig:atomic_dark_matter_f_vs_mmin} shows $f_{\text{DBH},90\%}(M_\text{min})$ derived from $\langle VT \rangle_{\rm O4a}$ and from the combined $\langle VT \rangle_{\rm O4a+b}$ for \GSTLAL, \MBTA, and \PYCBC individually.
As described in Section~\ref{sec:3-search_sensitivity}, the combined $\langle VT \rangle_{\rm O4a+b} > \langle VT \rangle_{{\rm O4}j}$ where $j={\rm a, b}$.
Hence, the combined analysis provides the most stringent constraint.

Applied to $\langle VT \rangle_{\rm O4a}$ alone, the 90\% upper limit reaches $f_{\rm DBH} = \DBHConstraintOfouraBest$, where the range arises from the differing sensitivities of the search pipelines.
Including \ac{O4b} tightens this by nearly a factor of \DBHConstraintFactorOfouraToOfourb, with the combined $\langle VT \rangle_{\rm O4a+b}$ giving $f_{\rm DBH} = \DBHConstraintOfourbBest$.
These are reached at $M_\text{min} \approx \DBHConstraintBestMminOfoura$ and $\DBHConstraintBestMminOfourb$ for \ac{O4a} and \Ofourab, respectively.
These limits on $f_\text{DBH}$ are consistent with the dissipative dark matter scenario as the formation channel requires efficient cooling, which only occurs in regions dense in dark matter.
We exclude limits where $M_\text{min} > 1\,\Msun$ as an \ac{SSM} detection necessitates the minimum mass to be smaller than $1\,\Msun$.
Conditioning instead on $f_\text{DBH}$ and integrating over $M_\text{min}$ gives the complementary statement that for $f_\text{DBH} = 1$ the analysis excludes $M_\text{min} > \DBHMminOfouraBound$ using $\langle VT \rangle_{\rm O4a}$ alone, tightening to $M_\text{min} > \DBHMminOfourbBound$ with the combined $\langle VT \rangle_{\rm O4a+b}$.
These constraints appear weaker than those of~\citet{LIGOScientific:2026wxz}, where the exclusion region was reported to extend as low as $f_{\text{DBH}} = \fconstraintOfouraold$ at $1\,\Msun$ and $M_{\min} > \mminlowerboundold$ at $f_{\text{DBH}} = 1$, respectively.
Those bounds were obtained from a global two-dimensional 90\% highest posterior density region, considering the joint posterior across the full $(M_{\min},f_{\mathrm{DBH}})$ parameter space.
That construction introduces a non-local dependence on both parameters: the location of the 90\% contour in a given region of parameter space depends on how strongly the posterior is constrained elsewhere in the joint distribution. 
(For example the quoted bound on $M_{\min}$ at $f_{\mathrm{DBH}} = 1$ depends on how strongly the population is constrained near $M_{\min} = 1\,\Msun$.) 
In contrast the bounds we quote here are conditional on a given value of $M_{\min}$ or of $f_{\mathrm{DBH}}$, hence are independent of how the posterior behaves elsewhere. 

The range of dark fermion masses $m_\chi$ probed by this search is set by $M_{\rm min}$ via dark Chandrasekhar mass, $M_\text{Chandra}^\text{Dark} \propto M_\text{Planck}^3/(\mu^\text{Dark}m_\chi)^2 \leq M_\text{min},$ where $\mu^\text{Dark} m_\chi$ is the mass per degenerate fermion~\citep{Chandrasekhar:1931ih}.
We derive a bound on the ratio of $m_\chi$ to the mass of the Standard Model proton $m_p$ by taking the ratio of the stellar Chandrasekhar mass to the dark Chandrasekhar mass as
\begin{equation}
    \mu^\text{Dark}\frac{m_\chi}{m_p} = 2\sqrt{\frac{1.4\,\Msun}{M_\text{Chandra}^\text{Dark}}} \geq 2 \sqrt{\frac{1.4\,\Msun}{M_\text{min}}}\,,
\end{equation}
where we set the stellar Chandrasekhar mass to $1.4\,\Msun$ and the corresponding $\mu^\text{Stellar}$ to $2$.

Previous searches~\citep{2023MNRAS.524.5984L} used the Standard Model value of $\mu^\text{Dark}=2$, but we report the bound leaving $\mu^\text{Dark}$ unspecified to cover a broader range of dissipative dark matter models, as done in~\citet{LIGOScientific:2026wxz}.
The lower end of the range is $\darkprotonlower$, set by the largest minimum mass considered, $M_\text{min} = 1\,\Msun$, and is common to both analyses.
Using $\langle VT \rangle_{\rm O4a}$ alone, the search probed $\mu^\text{Dark}m_\chi/m_p$ up to $\darkprotonOfouraupper$, extending to $\darkprotonOfourbupper$ with the combined $\langle VT \rangle_{\rm O4a+b}$, where the spread arises from the differing sensitivities of the search pipelines.
Again, these bounds appear weaker than those reported in~\citet{LIGOScientific:2026wxz}, which reported to probe $\mu^\text{Dark}m_\chi/m_p$ up to $\darkprotonupperold$, because the bound is obtained directly from the corresponding constraint on $M_{\min}$.
A future detection would be able to put a direct bound on the microphysics of dissipative dark matter.

%% file: source__6-conclusion.tex
\section{Conclusion}\label{6-conclusion}

We present results from searches for \ac{SSM} compact objects in data of the 
\acl{O4b} of the \acl{LVK}.
No significant candidate is reported by any of the three participating search 
pipelines. Outlier candidates identified by the \GSTLAL pipeline and
the \MBTA pipeline do not reach a sufficient level of significance to provide
evidence for detection, and are consistent with noise fluctuations. 
Given the null detection result, we report updated 
upper limits on the merger rates for \ac{SSM} \acp{BBH} with masses greater than 
$0.2\,\Msun$ using the estimated search sensitivity.

Using the inferred sensitive hypervolume and limits on the merger rate of \ac{SSM} 
\acp{BBH}, we update the constraints on two dark matter models: \acp{PBH}, and 
\acp{DBH} that form in a dissipative dark matter model.
While both models can populate the sub-solar and super-solar regimes, we turn to the 
\ac{SSM} parameter space for clear evidence of such objects because no known 
astrophysical channel produces \ac{SSM} black holes.

In the \ac{PBH} scenario, we set informative constraints on the fraction of 
dark matter in \acp{PBH} $f_\mathrm{PBH}$ in late-forming binaries, as done 
previously in~\citet{LIGOScientific:2026wxz}: 
we infer that $f_\mathrm{PBH} < 1$ at the 90\% confidence interval for 
$m_\mathrm{PBH} \gtrsim \LBfPBHBelowOneOfourAB$. 
For early-forming \ac{PBH} binaries, 
adopting a conservative model that accounts for binary disruption, we 
obtain $f_\mathrm{PBH}$ upper limits between \EBfPBHMinMPBHOfourAB and 
\EBfPBHMaxMPBHOfourAB over the
range of \ac{PBH} masses probed by  
the searches for \ac{SSM} systems in \ac{O4a} and \ac{O4b}.
For the dissipative dark matter scenario, where \ac{SSM} \ac{DBH} are assumed to form 
with some minimum possible mass, we bound the fraction of the total dark matter density 
in the Universe in \acp{DBH}; our most restrictive limit, for a minimum mass 
\DBHConstraintBestMminOfourb, is \DBHConstraintOfourbBest.

The \ac{SSM} searches using \ac{O4a} data provided  
the sensitivity of \ac{LVK} \ac{SSM} searches to binaries containing low-mass neutron 
stars for the first time by including tides in the simulated signals. 
In this paper, we updated the sensitivity for this region using our \ac{SSM} analyses 
of \ac{O4b} data. 
We find the typical sensitive hypervolume to be 
$\langle VT \rangle$ \BNSVTorderofmagnitude for populations of 
\acp{BNS} with at least one \ac{SSM} neutron star.
These searches can also provide insight into matter at high densities given the 
overlap of the neutron star and \ac{SSM} parameter space, serving as motivation 
for low-latency \ac{SSM} searches~\citep{2025CQGra..42j5009A, Hanna:2024tom} which 
started to release alerts 
on March 4, 2025, and continued through the end of \ac{O4}. 

The LIGO and Virgo detectors continued observations in the \ac{O4c} with similar 
sensitivities to \ac{O4b}, and were joined by the KAGRA detector for parts of 
the remainder of \ac{O4}. 
The increase in observing time is expected to increase our search sensitivities, 
leading to improvements to the constraints we can place on the merger rates of 
\ac{SSM} \acp{BBH}.
\Ac{O4} ended on November 18, 2025, and the next data collection period is 
planned to begin in late 2026, with subsequent runs expected to approach the design 
sensitivity of the detector network.\footnote{Current observing plans are 
maintained at \url{https://observing.docs.ligo.org/plan/}.}
These advances will provide more stringent limits on the fraction of dark matter 
in \ac{SSM} compact objects, or possibly detect a \ac{GW} signal 
originating from the merger of \ac{SSM} compact objects.
Such a detection would have deep implications for fundamental physics.

%% file: data_availability.tex
All strain data analysed as part of this publication are publicly available through \ac{GWOSC}.
The details of this data release are described in detail in~\citet{LIGOScientific:2026jgl}.
Associated data release includes trigger files for candidates reported in Table~\ref{tab:triggers}, \ac{BBH} sensitive volume-time, and injection results used in the calculation of sensitive volume-time reported in this paper~\citep{ZenodoO4bSSM}.

%% file: LVK_P2000488_v31.tex
This material is based upon work supported by NSF's LIGO Laboratory, which is a
major facility fully funded by the National Science Foundation.
The authors also gratefully acknowledge the support of
the Science and Technology Facilities Council (STFC) of the
United Kingdom, the Max-Planck-Society (MPS), and the State of
Niedersachsen/Germany for support of the construction of Advanced LIGO 
and construction and operation of the GEO\,600 detector. 
Additional support for Advanced LIGO was provided by the Australian Research Council.
The authors gratefully acknowledge the Italian Istituto Nazionale di Fisica Nucleare (INFN),  
the French Centre National de la Recherche Scientifique (CNRS) and
the Netherlands Organization for Scientific Research (NWO)
for the construction and operation of the Virgo detector
and the creation and support  of the EGO consortium. 
The authors also gratefully acknowledge research support from these agencies as well as by 
the Council of Scientific and Industrial Research of India, 
the Department of Science and Technology, India,
the Science \& Engineering Research Board (SERB), India,
the Ministry of Human Resource Development, India,
the Spanish Agencia Estatal de Investigaci\'on (AEI),
the Spanish Ministerio de Ciencia, Innovaci\'on y Universidades,
the European Union NextGenerationEU/PRTR (PRTR-C17.I1),
the ICSC - CentroNazionale di Ricerca in High Performance Computing, Big Data
and Quantum Computing, funded by the European Union NextGenerationEU,
the Comunitat Auton\`oma de les Illes Balears through the Conselleria d'Educaci\'o i Universitats,
the Conselleria d'Innovaci\'o, Universitats, Ci\`encia i Societat Digital de la Generalitat Valenciana and
the CERCA Programme Generalitat de Catalunya, Spain,
the Polish National Agency for Academic Exchange,
the National Science Centre of Poland and the European Union - European Regional
Development Fund;
the Foundation for Polish Science (FNP),
the Polish Ministry of Science and Higher Education,
the Swiss National Science Foundation (SNSF),
the Russian Science Foundation,
the European Commission,
the European Social Funds (ESF),
the European Regional Development Funds (ERDF),
the Royal Society, 
the Scottish Funding Council, 
the Scottish Universities Physics Alliance, 
the Hungarian Scientific Research Fund (OTKA),
the French Lyon Institute of Origins (LIO),
the Belgian Fonds de la Recherche Scientifique (FRS-FNRS), 
Actions de Recherche Concert\'ees (ARC) and
Fonds Wetenschappelijk Onderzoek - Vlaanderen (FWO), Belgium,
the Paris \^{I}le-de-France Region, 
the National Research, Development and Innovation Office of Hungary (NKFIH), 
the National Research Foundation of Korea,
the Natural Sciences and Engineering Research Council of Canada (NSERC),
the Canadian Foundation for Innovation (CFI),
the Brazilian Ministry of Science, Technology, and Innovations,
the International Center for Theoretical Physics South American Institute for Fundamental Research (ICTP-SAIFR), 
the Research Grants Council of Hong Kong,
the National Natural Science Foundation of China (NSFC),
the Israel Science Foundation (ISF),
the US-Israel Binational Science Fund (BSF),
the Leverhulme Trust, 
the Research Corporation,
the National Science and Technology Council (NSTC), Taiwan,
the United States Department of Energy,
and
the Kavli Foundation.
The authors gratefully acknowledge the support of the NSF, STFC, INFN and CNRS for provision of computational resources.

This work was supported by MEXT,
the JSPS Leading-edge Research Infrastructure Program,
JSPS Grant-in-Aid for Specially Promoted Research 26000005,
JSPS Grant-in-Aid for Scientific Research on Innovative Areas 2402: 24103006,
24103005, and 2905: JP17H06358, JP17H06361 and JP17H06364,
JSPS Core-to-Core Program A.\ Advanced Research Networks,
JSPS Grants-in-Aid for Scientific Research (S) 17H06133, 20H05639 and 26K21721,
JSPS Grant-in-Aid for Transformative Research Areas (A) 26A204: JP26H00407,
the joint research program of the Institute for Cosmic Ray Research,
University of Tokyo,
the National Research Foundation (NRF),
the Computing Infrastructure Project of the Global Science experimental Data hub
Center (GSDC) at KISTI,
the Korea Astronomy and Space Science Institute (KASI),
the Ministry of Science and ICT (MSIT) in Korea,
Academia Sinica (AS),
the AS Grid Center (ASGC) and the National Science and Technology Council (NSTC)
in Taiwan under grants including the Science Vanguard Research Program,
the Advanced Technology Center (ATC) of NAOJ, 
the Mechanical Engineering Center of KEK
and Vietnam National Foundation for Science and Technology Development 
(NAFOSTED) 103.01-2025.147.

Additional acknowledgements for support of individual authors may be found in the following document: \\
\texttt{https://dcc.ligo.org/LIGO-M2300033/public}.
For the purpose of open access, the authors have applied a Creative Commons Attribution (CC BY)
license to any Author Accepted Manuscript version arising.
We request that citations to this article use 'A. G. Abac {\it et al.} (LIGO-Virgo-KAGRA Collaboration), ...' or similar phrasing, depending on journal convention.

%% file: software_ack.tex
Calibration of the \ac{LIGO} strain data was performed with a \GSTLAL{}-based
calibration software pipeline~\citep{2018CQGra..35i5015V}.
Data-quality products and event-validation results were computed using the
\textsc{DMT}{}~\citep{DMTdocumentation}, \textsc{DQR}{}~\citep{DQRdocumentation},
\textsc{DQSEGDB}{}~\citep{2021SoftX..1400677F}, \textsc{gwdetchar}{}~\citep{gwdetchar-software},
\textsc{hveto}{}~\citep{2011CQGra..28w5005S}, \textsc{iDQ}{}~\citep{Essick:2020qpo},
\textsc{Omicron}{}~\citep{2020SoftX..1200620R}, \textsc{PEMCHECK}~\citep{2024CQGra..41n5003H} and
\textsc{PythonVirgoTools}{}~\citep{pythonvirgotools} software packages and contributing
software tools.
Analyses relied upon the \LALSUITE{} software
library~\citep{lalsuite, 2020SoftX..1200634W}.
The identification of the signal candidates and subsequent
significance evaluations in this paper were performed with the
\GSTLAL{}-based inspiral software
pipeline~\citep{2017PhRvD..95d2001M,2019arXiv190108580S,2020PhRvD.101b2003H,2021SoftX..1400680C},
with the \MBTA{} pipeline~\citep{2016CQGra..33q5012A,2021CQGra..38i5004A}, and with the
\PYCBC{} pipeline~\citep{2016CQGra..33u5004U,2017ApJ...849..118N,2020PhRvD.102b2004D}.
Sensitive Hypervolume calculations were performed using \GWPOPULATION~\citep{Talbot2025}.
Plots were prepared with \MATPLOTLIB{}~\citep{2007CSE.....9...90H},
\SEABORN{}~\citep{2021JOSS....6.3021W} and \GWPY{}~\citep{gwpy-software}.
\NUMPY{}~\citep{2020Natur.585..357H}, \SCIPY{}~\citep{2020NaMet..17..261V} and \PANDAS{}~\citep{the_pandas_development_team_2024_10537285} were used
in the preparation of the manuscript.